\documentclass[10pt,twocolumn,superscriptaddress,english,10pt,showpacs,floatfix,prl]{revtex4-1}
\usepackage{amsthm}
\usepackage{amsmath}
\usepackage{amssymb}
\usepackage{mathdots}
\usepackage{graphicx}
\usepackage{wasysym}
\usepackage{xspace}

\usepackage[normalem]{ulem}
\usepackage{array}
\makeatletter
\@ifundefined{textcolor}{}
{%
 \definecolor{BLACK}{gray}{0}
 \definecolor{WHITE}{gray}{1}
 \definecolor{RED}{rgb}{1,0,0}
 \definecolor{GREEN}{rgb}{0,1,0}
 \definecolor{BLUE}{rgb}{0,0,1}
 \definecolor{CYAN}{cmyk}{1,0,0,0}
 \definecolor{MAGENTA}{cmyk}{0,1,0,0}
 \definecolor{YELLOW}{cmyk}{0,0,1,0}
}

\usepackage{braket}
\newcommand{\gv}[1]{\ensuremath{\mbox{\boldmath$ #1 $}}} 
\newcommand{\abs}[1]{\left| #1 \right|} % for absolute value
\renewcommand{\d}[2]{\frac{d #1}{d #2}} % for derivatives
\let\baraccent=\= % rename builtin command \= to \baraccent
\renewcommand{\=}[1]{\stackrel{#1}{=}} % for putting numbers above =

\newcolumntype{L}[1]{>{\raggedright\let\newline\\\arraybackslash\hspace{0pt}}m{#1}}
\newcolumntype{C}[1]{>{\centering\let\newline\\\arraybackslash\hspace{0pt}}m{#1}}
\newcolumntype{R}[1]{>{\raggedleft\let\newline\\\arraybackslash\hspace{0pt}}m{#1}}

\makeatother

\usepackage{babel}

\begin{document}
\title{Time-quasiperiodic topological superconductors with Majorana Multiplexing}
\author{Yang Peng}
\email{yangpeng@caltech.edu}
\affiliation{Institute of Quantum Information and Matter and Department of Physics,California Institute of Technology, Pasadena, CA 91125, USA}
\affiliation{Walter Burke Institute for Theoretical Physics, California Institute of Technology, Pasadena, CA 91125, USA}
\author{Gil Refael}
\affiliation{Institute of Quantum Information and Matter and Department of Physics,California Institute of Technology, Pasadena, CA 91125, USA}

\begin{abstract}
	Time-quasiperiodic Majoranas are generalizations of Floquet Majoranas in time-quasiperiodic superconducting systems. We show that in a system driven at $d$ mutually
	irrational frequencies, there are up to $2^d$ types of such Majoranas, coexisting despite spatial overlap and lack of time-translational invariance. 	Although the
	quasienergy spectrum is dense in such systems,  the time-quasiperiodic Majoranas can be stable and robust against resonances due to localization in the periodic-drives
	induced synthetic dimensions. This is demonstrated in a time-quasiperiodic Kitaev chain driven at two frequencies. We further relate the existence of multiple Majoranas in a time-quasiperiodic system to the time quasicrystal phase introduced recently. These time-quasiperiodic Majoranas open a new possibility for braiding which will be pursued in the future.
\end{abstract}

\maketitle

{\em Introduction.---} 
Majorana bound states, aka Majoranas, are zero-energy excitations in topological superconductors
ninvariant under particle-hole transformation \cite{Kitaev2001,Alicea2012,Beenakker2013}. 
Their zero-energy nature gives rise to degenerate ground states, 
which can be used as nonlocal qubits and memory \cite{Kitaev2003,Nayak2008,Aasen2016}.
Therefore, Majorana engineering in a variety of platforms has been an simmering field of study
both theoretically
\cite{Fu2008,Zhang2008,Sato2009,Lutchyn2010,Oreg2010,Diehl2011,Jiang2011,Nadj-Perge2013,Pientka2013,Foster2014,Peng2015}
and experimentally
\cite{Mourik2012,Das2012,Churchill2013,Deng2012,Finck2013,Nadj-Perge2014,Ruby2015,Pawlak2016,Deng2016,Albrecht2016,Ruby2017,Gul2018}.

Topological phases, however, also exist under nonequiliubrium conditions and
can be realized by time-periodic driving, known as Floquet engineering. 
%In particular, it was predicted that a static band insulator
%can be brought into a topological phase
%by circularly polarized radiation or an alternating Zeeman field
%\cite{Oka2009,Inoue2010,Kitagawa2011,Lindner2011,Lindner2013}.
Floquet topological superconductors and superfluids were proposed to be realized in either periodically
driven cold atom systems \cite{Jiang2011} or proximitzed nanowires \cite{Klinovaja2016,Thakurathi2017}.
Floquet topological phases have also been explored 
experimentally \cite{Wang2013,Jotzu2014,Aidelsburger2015,Tarnowski2017,Maczewsky2017}.

Interestingly, Floquet topological superconductors (or superfluids) 
host a dynamical version of Majoranas, dubbed Floquet Majoranas \cite{Jiang2011,Liu2013}.
Rather than sitting at zero enregy, 
Floquet Majoranas have quasienergies $\epsilon=0$ or $\epsilon=\omega/2$, 
where $\omega$ is the driving frequency. Because energy is only defined modulo $\omega$, $\omega/2$ is a particle-hole symmetric point in the spectrum just as $\epsilon=0$ is, and the particle-hole symmetric nature
of these Majoranas holds in a time-dependent fashion at all times. 
Indeed, Floquet Majoranas can form topological qubits and store quantum information,
just as their equilibrium counterparts do \cite{Liu2013}. 
Floquet Majoranas may therefore open a new route for topological quantum computation using the time domain as a resource \cite{time-braiding}.

\begin{figure}[t]
	\centering
	\includegraphics[width=0.4\textwidth]{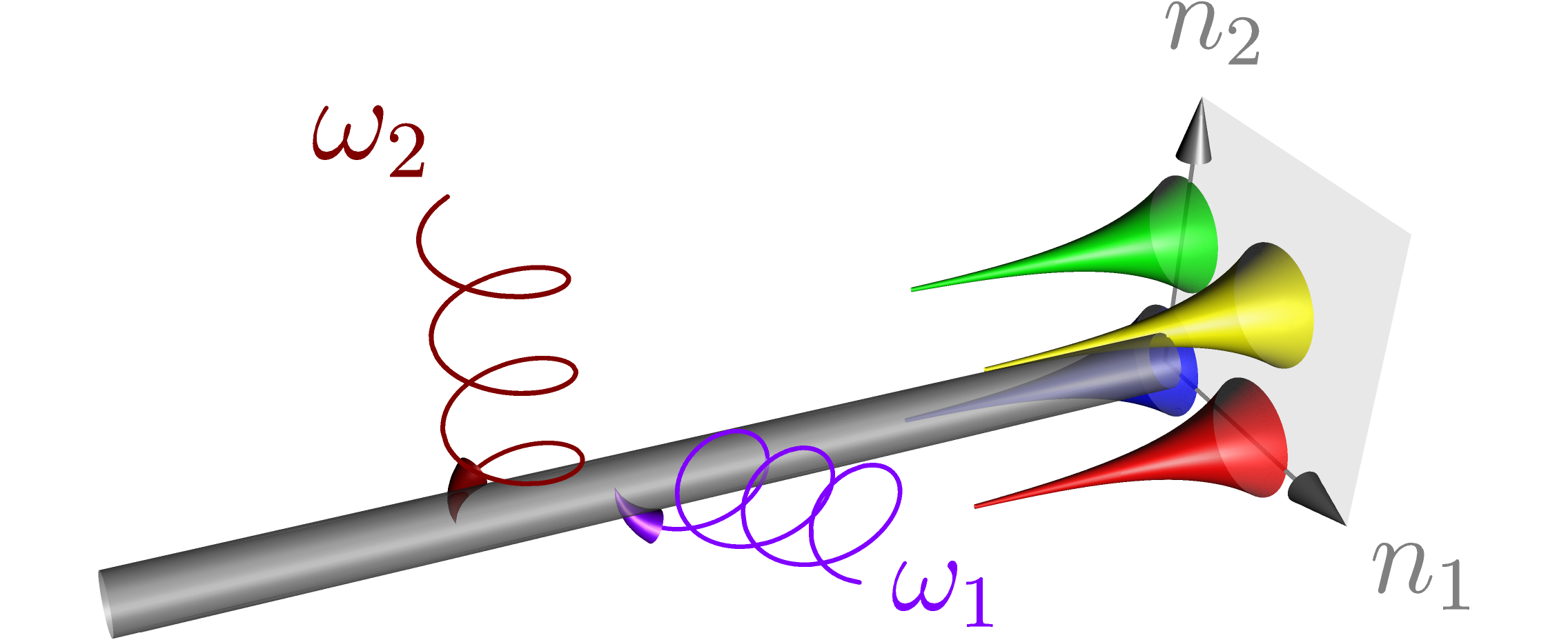}
	\caption{\label{fig:wire} Schematic representation of time-quasiperiodic Majoranas localized at the end of a
  1D topological superconductor (in grey) driven at two frequencies $\omega_1$ and $\omega_2$. These Majoranas are localized
  in both real space and the two synthetic dimensions with coordinates $n_1$ and $n_2$. }
\end{figure}

A natural question arises: could topological behavior also arise when a drive contains multiple frequencies, without any time-translational invariance? If so, could we obtain multiple Majorana modes associated with these frequencies? This would be similar to frequency multiplexing to enhance
the hardware channel capacity in optical fibers \cite{Tomlinson1977}.
For concreteness, let us consider a time-quasiperiodic superconductor driven at two frequencies $\omega_1$ and $\omega_2$, 
where $\omega_1/\omega_2$ is an irrational number, otherwise the system is time-periodic.
We assume the concept of quasienergy (as we will introduce it later) also exist in this context, 
which is defined up to $n_1\omega_1+n_2\omega_2$ with $n_1,n_2 \in \mathbb{Z}$. 
Thus, there are four inequivalent particle-hole symmetric quasienergies: $0$, $\omega_1/2$, $\omega_2/2$, and
$(\omega_1+\omega_2)/2$. This means one can at most have four types of Majoranas, as
shown in Fig.~\ref{fig:wire}. On the other hand, from a naive point of view, since $n_1\omega_1+n_2\omega_2$
could be made to yield arbitrary energy increments, as long as $\abs{n_1},\abs{n_2}$ are large enough,  
the quasienergy spectrum will be everywhere dense, with multi-photon energy arbitrarily small near resonances, and these Majoranas appear fully unstable. 

In this manuscript, we demonstrate that multi-frequency driven systems can give rise to a new class of time-quasiperiodic topological phases. Furthermore, such time-quasiperiodic topological superconductors give rise to Majorana edge states appearing at several frequencies simultaneously. These multiple Majoranas are stable and can coexist due to localization in the drive-induced synthetic $n_1$ and $n_2$ dimensions, which also suppresses the hybridization between the Majorana edge states, and bulk extended states. This renders the Majorana edge modes as  stable spatially localized edge states.  We confirm this by simulating a Kitaev chain driven at two incommensurate frequencies,  and show the existence of Majorana edge states with half-frequency quasienergies. Furthermore, we use our simulations to demonstrate that time-quasiperiodic Majoranas are related to 
the ``time quasicrystal'' phases introduced recently in
time-quasiperiodic spin chains \cite{Dumitrescu2018} (see also Refs.~\cite{Li2012,Flicker2017, Autti2018,Huang2018,Giergiel2018}); the half-frequency Majoranas are essentially the single-particle degrees of freedom  characterizing the time-quasicrystal phase, in the same vain that the
Floquet Majoranas are underlying the time-crystal period doubling of Refs.~\cite{Khemani2016,Else2016,Potter2016,Bomantara2018}.

{\em Floquet recap---} 
Let us start by briefly reviewing Floquet states. Consider a time-periodic Hamiltonian $H(t) = H(t+T)$, with
driving angular frequency $\omega$, and  period $T=2\pi/\omega$.
The solutions to the time-dependent Schr\"odinger equation are characterized by the Floquet states, given by $\ket{\Psi_{\alpha}(t)} = e^{-i\epsilon_\alpha t} \ket{\Phi_{\alpha}(t)}$, 
where $\ket{\Phi_{\alpha}(t)}$ is a periodic function with the same period as the Hamiltonian, which satisfies the eigenvalue equation $[H(t) - i\partial_t] \ket{\Phi_\alpha(t)} = \epsilon_\alpha \ket{\Phi_\alpha(t)}$
with eigenvalue $\epsilon_{\alpha}$. Here, $K(t) = H(t) - i\partial_t$ and $\epsilon_\alpha$ are called quasienergy operator and quasienergy, respectively. 

It is important to note that quasienergies are not uniquely defined. Indeed, $\epsilon_\alpha$ and
$\epsilon_{\alpha,n}=\epsilon_\alpha+n\omega$ with $n\in \mathbb{Z}$ actually describe the same physical state
$\ket{\Psi_\alpha(t)}=e^{-i\epsilon_\alpha t}\ket{\Phi_\alpha(t)}=e^{-i\epsilon_{\alpha,n}
t}\ket{\Phi_{\alpha,n}(t)}$, 
where $\ket{\Phi_{\alpha,n}(t)} = e^{in \omega t}\ket{\Phi_\alpha(t)}$ is also an eigenfunction 
of the quasienergy operator at quasienergy $\epsilon_{\alpha,n}$. 
Thus, the quasienergy $\epsilon_\alpha$ is only uniquely defined modulo $\omega$, e.g., in the range $-\omega/2\le\epsilon<\omega/2$.

{\em Floquet synthetic dimensions and Wannier-Stark localization---}
Our construction of time-quasiperiodic Majoranas requires recasting the driven Hamiltonian in a time-independent way.  Let us write out the Hamiltonian and Floquet states using their
Fourier expansion of $H(t) = \sum_{n}e^{-im \omega t} h_{n}$ and $\ket{\Phi_{\alpha}(t)} =\sum_{m}e^{-im\omega
t}\phi_m^\alpha$.
The eigenvalue equation for the quasienergies then becomes
\begin{equation}
\sum_{m}h_{n-m}\phi_{m}^{\alpha} - n\omega\phi_{n}^{\alpha}
= \epsilon_\alpha \phi_{n}^\alpha, \label{synH}
\end{equation}
which describes particles hopping in a 1D synthetic lattice, spanned by the coordinate $n$, with $\omega$ playing the role of a uniform force field. This is precisely the Hamiltonian for a Wannier-Stark
ladder, with energy difference $\omega$ between neighboring rungs.
We will restrict ourselves to nearest-neighbor-hopping models, i.e. $h_{n}=0$ for $\abs{n}\geq 2$. 

It has been known that the electronic wave functions in the Wannier-Stark ladder are localized, 
with a localization length $\sim 1/\ln(\omega/V)$ when $V<\omega$, with $V$ being the nearest neighbor hopping amplitude, known as the Wannier-Stark localization \cite{Fukuyama1973,Emin1987}. Likewise we expect that the Floquet states will be localized to the vicinity of a particular $n$, which is a manifestation of energy conservation. 

{\em Floquet Particle-hole symmetry in superconductors.}
The hamiltonians of superconductors possess a unitary matrix $U_{P}$
such that $U_{P}H(t)^{*}=-H(t)U_{P}$ for all times, with ``$*$'' denoting complex conjugation. 
This particle-hole symmetry dictates that $U_{P}K(t)^{*}U_{P}^{\dagger}=-K(t)$, and that the Floquet states appear in pairs as $\ket{\Phi_\alpha(t)}$ and $U_{P}\ket{\Phi_\alpha(t)^*}$,
with quasienergies $\pm\epsilon_\alpha$, respectively.

Majoranas are special states that are particle-hole symmetric. Namely,  with $\ket{\psi(t)}$ a Majorana state:
\begin{equation}
 e^{-i\varepsilon t}\ket{\phi(t)}= \ket{\psi(t)}=U_P \ket{\psi(t)^*}= e^{i\varepsilon t}U_P\ket{\phi(t)^*},
\end{equation}
which works if $(U_P\ket{\phi(t)^*}) = e^{-ip\omega} \ket{\phi(t)}=e^{-2i\varepsilon t}\ket{\phi(t)}$ with some $p\in\mathbb{Z}$. Therefore, the majorana quasienergies are restricted to $\varepsilon = p\omega/2$ with some $p\in\mathbb{Z}$.
And because shifts by $\omega$ are just a gauge choice, there are only two inequivalent Floquet Majoranas \cite{Jiang2011,Liu2013}, with $p\in \{0,1\}$ reduced to a $\mathbb{Z}_2$ variable.

{\em Floquet Majoranas.} Next consider a 1D Floquet topological superconductor, with Hamiltonian  $H(t) = H_{K} + M(\omega t)$. The first term describes a static Kitaev chain
\begin{equation}
	H_{K} = -\mu\sum_{j=1}^N c_{j}^\dagger c_{j} -\sum_{j=1}^{N-1}[(Jc_{j}^\dagger c_{j+1} + i\Delta c_{j} c_{j+1}) + h.c.].
  \label{eq:kitaev}
\end{equation}
with  $c_j$ ($c_j^\dagger$) annihilation (creation) operators at site $j$, $\mu$ is the chemical potential, $J$ is the hopping amplitude, 
and $\Delta$ is the $p$-wave pairing potential.
The second term, 
\begin{equation}
	M(\omega t) = -i\Delta' \sum_{j=1}^{N-1}( e^{-i\omega t} c_{j} c_{j+1} - e^{i\omega t}c_{j+1}^\dagger c_{j}^\dagger),
  \label{eq:M_pairing}
\end{equation}
corresponds to a periodic drive.  Introducing Nambu spinors in momentum ($k$) space $\Psi_k^\dagger = (c_k^\dagger, c_{-k})$, with $c_k = \sum_{j=1}^N c_j e^{-i k j}/\sqrt{N}$. For periodic boundary conditions,  we get the Bogoliubov--de Gennes Hamiltonian
\begin{equation}
\begin{array}{c}
H=\sum_{k>0}\Psi_k^\dagger[\mathcal{H}_K(k)+\mathcal{M}(k,\omega t)]\Psi(k),\vspace{2mm}\\
\mathcal{H}_K(k) = \tau_z\xi_k +\tau_x\Delta\sin k ,\,\, \mathcal{M}(k,\omega t)=\tau_x \Delta' \sin k e^{i\omega t\tau_z}
\end{array}
\end{equation}
 where $\tau_{x,y,z}$ are the Pauli matrices in
Nambu space, and $\xi_k = -J\cos k -\mu/2$ is the normal state dispersion. 

\begin{figure}[t]
	\centering
	\includegraphics[width=0.48\textwidth]{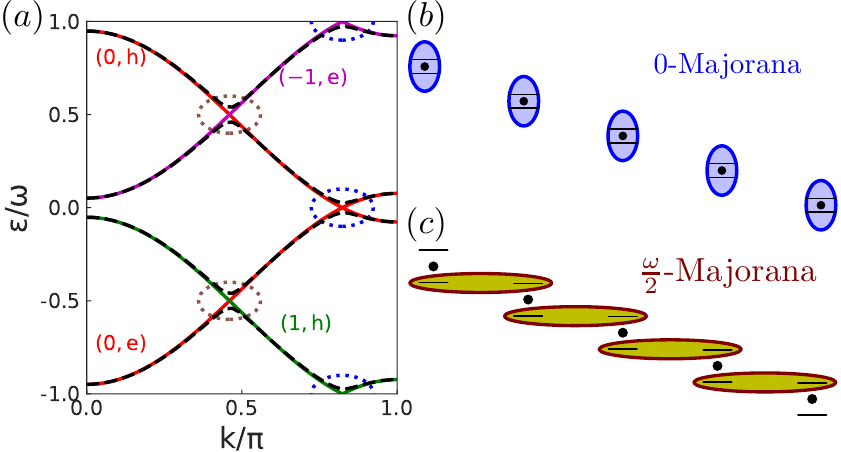}
	\caption{\label{fig:floquet-ladder} $(a)$ Quasienergy spectrum as a function of $k$ between $-\omega$ and $\omega$, for the model defined in
		Eqs.~(\ref{eq:kitaev},\ref{eq:M_pairing}). The black dashed lines are obtained with $J/\omega=0.51$, $\mu/\omega=0.87$,  $\Delta/\omega=0.051$, and
		$\Delta'/\omega=0.038$.
  The solid red, green, and magenta lines corresponding to the quasienergies $\epsilon_{n,e/h}$ when setting
  $\Delta=\Delta'=0$, for a certain $(n,e/h)$, as indicated in the figure with the same color. The two types of
  topological gaps are indicated in the blue and brown dotted circles.  $(b)$ and $(c)$ are the Wannier-Stark ladders
  with two orbitals (black lines) per rung (black dot),  when
  $\epsilon_{n,e}\simeq \epsilon_{n,h}$ and $\epsilon_{n+1,h}\simeq \epsilon_{n,e}$  respectively.
  The $0$-Majoranas are formed from equal superposition between states $(n,e)$ and $(n,h)$ (blue ellipses), 
  while the $\omega/2$-Majoranas are formed from equal superposition between states $(n+1,h)$ and $(n,e)$(green ellipses).}
\end{figure}

The spectrum of the driven Kitaev model can be interpreted using the synthetic dimension and Wannier-Stark-ladder approach of Eq. (\ref{synH}). For each $k$ there are two orbitals for each harmonic $n$.
Thus, in the absence of pairing potential, the system has two groups of
equally-spaced spectra $\epsilon_{n,e/h}= \pm \xi_k + n\omega$, with $n\in\mathbb{Z}$.
The $+$ or $-$ signs indicate electron-like (e) or hole-like (h) states. The static pairing potential $\Delta$ opens a topological
gap at $n\omega$, when $\epsilon_{n,e}=\epsilon_{n,h}$, while the  dynamical pairing $\Delta'$ opens a topological
gap at $(n+1)\omega/2$ when $\epsilon_{n+1,h}=\epsilon_{n,e}$, i.e., at the edge of the `Floquet zone'.
In Fig.~\ref{fig:floquet-ladder}$(a)$, we show the spectrum of the ladder
as a function of $k$ in a window between $-\omega$ and
$\omega$,  with a set of parameters producing the two topological gaps.
An open chain, then, supports two types of Floquet Majoranas at quasienergies $0,\,\omega/2$,
with same-rung  equal superposition of electron and hole states  (Fig.~\ref{fig:floquet-ladder}$(b)$), and between neighboring rungs (see Fig.~\ref{fig:floquet-ladder}$(c)$),
respectively.

{\em Time-quasiperiodic Majoranas.---}
Our main result is that Majoranas also emerge due to multi-frequency drive. Consider a time-quasiperiodic Hamiltonian
$H(t)$ characterized by $d$ mutually irrational frequencies $\boldsymbol{\omega}
=(\omega_1,\dots,\omega_d)$. The Floquet ansatz introduced previously can be generalized to the time-quasiperiodic
system \cite{suppl}. The function $\ket{\Phi_\alpha(t)}$,  which becomes time-quasiperiodic 
at frequencies specified by $\boldsymbol{\omega}$,
satisfies the eigenvalue equation of the time-quasiperiodic quasienergy operator $K(t)$:
\begin{equation}
K(t)\ket{\Phi_\alpha(t)}=\left(H(t) -i\frac{\partial}{\partial t}\right)\ket{\Phi_\alpha(t)}=\epsilon_\alpha \ket{\Phi_\alpha(t)}
\end{equation}
with the quasienergy $\epsilon_\alpha$ defined modulo $\mathbf{n}\cdot\boldsymbol{\omega}$.

Time-quasiperiodic Majoranas then emerge as particle-hole symmetric states. 
These must have quasienergies $\epsilon = \mathbf{p}\cdot\boldsymbol{\omega}/2$, 
with $\mathbf{p} \in \mathbb{Z}^d$. Furthermore, they fall into $2^d$ groups, 
reducing $\mathbf{p} \in \{0,1\}^d$, corresponding to $2^d$ types of Majoranas. 

Contrary to a gapped Floquet topological phase, the quasienergy spectra 
in a time-quasiperiodic system are dense, since 
$\mathbf{n}\cdot\boldsymbol{\omega}$ can approach any value.
It seems, therefore, that time-quasiperiodic Majoranas do not have a gap that could protect them from hybridizing with bulk states due to local perturbations. Below we show that these majoranas are stable not due to a gap, but rather due to localization in the drive-induced synthetic dimensions.

{\em Multidrive synthetic Lattice and localization---} 
Similar to the Floquet case, the time-quasiperiodic system 
could be posed as a time-independent problem. 
The quasienergy eigenvalue equation becomes a tight-binding
problem on a $d$-dimensional lattice whose coordinates are given by $\mathbf{n}\in \mathbb{Z}^d$ 
embedded in the $d$-dimensional Euclidean space $\mathbb{R}^{d}$.
In addition, a force field given by $\mathbf{\omega}$ pointing into the synthetic
dimensions keeps track of the energy of energy quanta absorbed from the drive \cite{Martin2017,Peng2018}. 

\begin{figure}[t]
	\centering
	\includegraphics[width=0.4\textwidth]{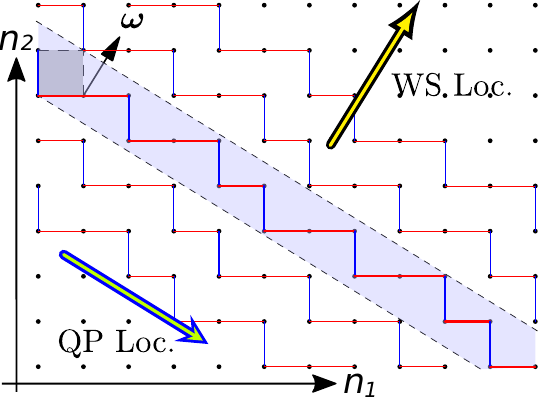}
	\caption{\label{fig:lattice}  2D synthetic lattice with an electric field vector
  $\boldsymbol{\omega}=(\omega_1,\omega_2)$ consisting of the driven frequencies. 
	The equipotential lines perpendicular to $\boldsymbol{\omega}$ are denoted as black dashed lines. 
	One obtains a 1D quasicrystal in between the two dashed lines denoted as the blue region. 
	The nearest-neighbor couplings within the quasicrstal
	are denoted as solid red or blue lines, corresponding to the original horizontal 
	and vertical couplings. The two big arrows denotes the directions along which
there are localizations: Wannier-Stark (WS) vs. Quasiperiodic (QP).}
\end{figure}

The equipotential surface perpendicular to the synthetic electric field defines a $(d-1)$-dimensional quasicrystal \cite{Duneau1985}. Fig.~\ref{fig:lattice} describes the quasicrystal construction for $d=2$, which is easily generalized to more dimensions. The lattice sites in a narrow strip (contained in the blue region) normal to the frequency vector $\boldsymbol{\omega}$ make a one-dimensional (1D) quasicrystal where the on-site energy goes up and down by $\omega_2$ and $\omega_1$. By shifting the strip along $\boldsymbol{\omega}$, the whole two-dimensional (2D) lattice will be covered, and every lattice sites
will be uniquely contained in one 1D quasicrystal.  Hence, the original system is equivalent to a Wannier-Stark ladder
of 1D quasicrystals.  Now it is clear, however, what can protect majoranas from bulk hybridization. Motion in a quasicrystal is fully localized if the hopping strength is smaller than the quasiperiodic modulation of the
on-site potential \cite{Aubry1980,Lahini2009}.

\begin{figure}[t]
	\centering
	\includegraphics[width=0.49\textwidth]{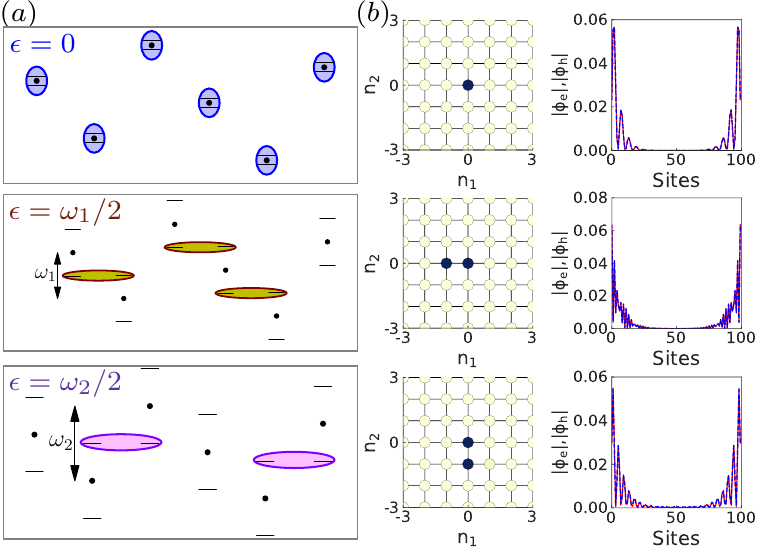}
	\caption{\label{fig:multiplexing} $(a)$The quasipeirodic ladder perpendicular 
		to $\boldsymbol{\omega}$ in the 2D synthetic lattice, with each rung corresponding
		to a Kitaev chain.  For a periodic chain, when $k$ is close to three special quasimomenta such that  $\epsilon_{n_1,n_2,e}\simeq \epsilon_{n_1,n_2,h}$ (top), 
  $\epsilon_{n_1+1,n_2,h}\simeq \epsilon_{n_1,n_2,e}$ (middle), and $\epsilon_{n_1,n_2+1,h}\simeq \epsilon_{n_1,n_2,e}$ (bottom), topological gaps 
  are induced. The three types of topological gaps give rise to three types of Majoranas in an open chain. 
  $(b)$ Numerical solution of the 0-frequency and time-quasiperiodic Majorana states on the 2D synthetic lattice of size $15 \times 15 $.
  Each site of the lattice corresponding to a Kitaev chain of length $N=100$.
  Left:  $\abs{\phi_{n_1,n_2}}^2$ for the $0$, $\frac{\omega_1}{2}$, and $\frac{\omega_2}{2}$  Majoranas
  on the 2D synthetic lattice, where the darker color corresponds to a larger magnitude. 
  Right: the absolute value of the corresponding Majorana wave function, summed over the 2D synthetic lattice.
  The electron and hole components $\phi_e, \phi_h$ are plotted as red solid and blue dashed curves.
  The other parameters are $\omega_2/\omega_1 =(\sqrt{5}+1)/2$, $J/\omega_1=0.51$, $\mu/\omega_1=0.87$, $\Delta/\omega_1=0.051$,
  and   $\Delta'/\omega_1=0.038$.
  }
\end{figure}

Therefore, Majoranas emerge from a combination of three localization mechanisms: 1) real space
localization due to the superconducting gap; 2) Wannier-Stark localization along the synthetic `electric' field,  $\boldsymbol{\omega}$  ;
3) Quasiperiodicity induced localization perpendicular to  $\boldsymbol{\omega}$. We focus on the
time-quasiperiodic Kitaev chain $H(t) = H_K + M(\omega_1 t) +M(\omega_2 t)$, following Eqs.~(\ref{eq:kitaev},
\ref{eq:M_pairing}), with $\frac{\omega_2}{\omega_1}=\frac{\sqrt{5}+1}{2}$. In the
synthetic space, $n_1,\,n_2$, of harmonics of the $\omega_1,\,\omega_2$ drives, the system is localized along the
$\boldsymbol{\omega}$ direction due to Wannier-Stark localization. The system is localized perpendicular to
$\boldsymbol{\omega}$ due to quasiperiodic localization when $\Delta'<\omega_1,\omega_2$. On a ring, there are two orbitals per rung for each quasimomentum $k$.
Ignoring the pairing potentials $\Delta,\Delta'$, 
the eigenvalues of this system are $\epsilon_{n_1,n_2,e/h} = \pm \xi_k +n_1\omega_1+n_2\omega_2$.
By choosing proper parameters, one has three special quasimomenta at which $\epsilon_{n_1,n_2,e}=\epsilon_{n_1,n_2,h}$, 
$\epsilon_{n_1+1,n_2,h}=\epsilon_{n_1,n_2,e}$, and $\epsilon_{n_1,n_2+1,h}=\epsilon_{n_1,n_2,e}$.
$\Delta$ and $\Delta'$, however, open topological gaps  at these crossings. In an open chain, these gaps give rise to three kinds of Majoranas, with quasienergies $0$, $\omega_1/2$ and $\omega_2/2$  (Fig.~\ref{fig:multiplexing}$(a)$).
The existence, stability, and localization of these Majoranas are verified via numerical simulation outlined in the
supplemental material~\cite{suppl}. Fig.~\ref{fig:multiplexing}$(b)$
shows these wavefunctions $\phi_{n_1,n_2} = (\phi_{n_1,n_2,e},\phi_{n_1,n_2,h})$ in the synthetic and real spaces.
Indeed, the wavefunction, which is identical for the hole and electron components, is localized at a single, or two
neighboring sites, in the synthetic directions, and near the edges in real space.

\begin{figure}[t]
	\centering
	\includegraphics[width=0.4\textwidth]{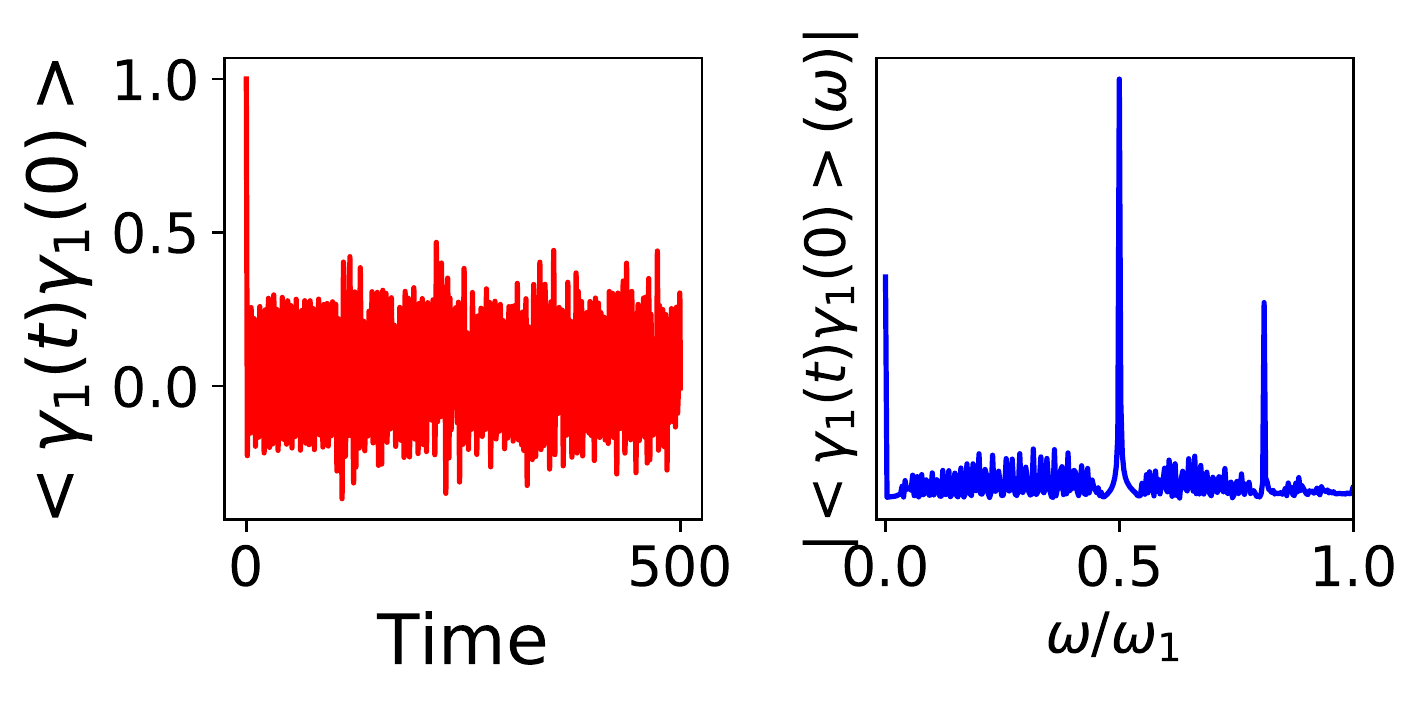}
	\caption{\label{fig:correlation} Left: Time evolution of $\langle \gamma_1(t) \gamma_1(0)\rangle$ simulated on the time-quasiperiodic Kitaev chain,  with 
	the same parameters as in Fig.~\ref{fig:multiplexing}. Right: The Fourier transform of $\langle \gamma_1(t) \gamma_1(0)\rangle$ in the frequency domain. 
	There are three dominant peaks at $0$, $\omega_1/2$, and $\omega_2/2 \simeq 0.81 \omega_1 $.
	}
\end{figure}

{\em From Majorana multiplexing to time quasicrystal.---}
The different types of Majoranas, gives rise to a quasiperiodic oscillating pattern distinct from the driving pattern
in the correlation function $\langle\hat{O}(t)\hat{O}(0)\rangle$ of a local observable
$\hat{O}$, resembling the time quasicrystal of Ref.~\cite{Dumitrescu2018}.
Take, for instance, $\hat{O}$ to be $\gamma_1 = (c_{1}+c_{1}^{\dagger})/\sqrt{2}$,
with $c_1,\, c_1^\dagger$ the electron creation and annihilation operators at the first site.
The correlation function is then closely related
to the local spectral function, and is dominated by the boundary modes, namely, the Majorana operators
\begin{equation}
\gamma_1(t)= c_0\psi_0(t)+c_1 \psi_1(t)+c_2 \psi_2(t)+\dots
\end{equation}
where $\psi_{0,1,2}$ are the time-quasiperiodic Majorana operators at quasienergies  $0$, $\omega_1/2$ and
$\omega_2/2$. Hence, $\langle\gamma_1(t)\gamma_1(0)\rangle$ generically contains peaks at frequencies, $0$, $\omega_1/2$ and $\omega_2/2$ (see
Fig.~\ref{fig:correlation}), where the average is with respect to the BCS vacuum at $t=0$.
In fact, the spectral peaks at half-frequencies persist even we include temporal disorders or take commensurate frequencies (see the Supp. Mat. Ref.~\cite{suppl} for details).

If one applies a Jordan-Wigner transform of the time-quasiperiodic Kitaev chain, we get a time-quasiperiodic Heisenberg
model.  $\langle \gamma_1(t) \gamma_1(0)\rangle$ becomes the spin correlation function $\langle \sigma_1^x(t)
\sigma_1(0)\rangle$.  This shows that the time-quasiperiodic Majoranas in a fermionic system are indeed the single-particle degrees of freedom which are responsible for the formation of the time quasicrystal correlations discussed in Ref.~\cite{Dumitrescu2018}.

{\em Conclusion. ---} In this work, we establish the existence of time-quasi-periodic topological phases, and generalize the concept of Floquet Majoranas to time-quasiperiodic systems.
We show that there are at most $2^d$ types of Majoranas at quasienergies $\mathbf{p}\cdot\boldsymbol{\omega}/2$,
with $\mathbf{p}\in \{0,1\}^d$ with $\boldsymbol{\omega}=(\omega_1,\dots,\omega_{d})$ consisting of
$d$ mutually irrational frequencies. Furthermore, we show that these Majorana states are stable, fully-localized, edge states. We study the time-quasiperiodic Kitaev chain with $d=2$,
and find coexisting stable and robust Majoranas at quasienergies $0$, $\omega_1/2$ and $\omega_2/2$. The localization in synthetic dimensions, emerges as a resource that allows  
these localized Majorana edge modes despite a dense quasienergy spectrum.
These Majoranas are also the single-particle degrees of freedom which are relevant to the formation of time quasicrystal \cite{Dumitrescu2018}.

The existence of time-quasiperiodic Majoranas opens a new direction for performing and controlling topological quantum computations using the time domain as a resource for topological anscilla qubits, for instance. Instead of using multiple static topological superconducting wires, one can dynamically generate multiple Majoranas at different locations for manipulation, by driving a single superconductor at different frequencies in different regions. While this raises issues of equilibration and heating, protocols for finite time manipulation may keep such problems at bay, even if these may be experimentally challenging at present.

{\em Acknowledgments.---}We acknowledge support from the IQIM,
an NSF physics frontier center funded in part by the Moore Foundation.
Y. P. is grateful to support from the Walter Burke Institute for Theoretical Physics at Caltech.
G. R. is grateful to support from the ARO MURI W911NF-16-1-0361 ``Quantum Materials by Design with Electromagnetic Excitation'' sponsored by the U.S. Army, as well as to the Aspen Center for Physics, supported by National Science Foundation grant PHY-1607761, where part of the work was done. "

\setcounter{equation}{0}
\begin{widetext}
\newpage
\section*{Supplemental Material}

\section*{Floquet ansatz for time-quasiperiodic systems}
In this section, we will prove the validity of Floquet ansatz in a time-quasiperiodic system.
Namely, the solution $\Psi(t)$ to a time-dependent Schr\"odinger equation in a time-quasiperiodic system can be written
as $\Psi(t)  = e^{-i\epsilon t}\Phi(t)$ with quasienergy $\epsilon$ and time-quasiperiodic $\Phi(t)$.

A time-dependent Hamiltonian $H(t)$ is time-quasiperiodic with 
$d$ frequencies if $H(t)=h(\omega_{1}t,\dots,\omega_{d}t)$, where $h(\theta_{1},\dots,\theta_{d})$
is a function of with $d$ $2\pi$-periodic arguments $\gv{\theta}=(\theta_{1},\dots,\theta_{d})$
living on a $d$-dimensional torus $\mathbb{T}^{d}=(\mathbb{R}/2\pi\mathbb{Z})^{d}$.
The frequencies $\gv{\omega}=(\omega_{1},\cdots,\omega_{d})$ are assumed
to be mutually irrational, namely
\begin{equation}
\sum_{j=1}^{d}n_{j}\omega_{j}\neq0,\quad\forall\mathbf{n}=(n_{1},\dots,n_{d})\in\mathbb{Z}^{d}.
\end{equation}

Consider the time evolution of an arbitrary state $\Psi(t)$ which
obeys the time-dependent Schr\"odinger equation (SEQ)
\begin{equation}
i\dot{\Psi}(t)=H(t)\Psi(t),\quad\dot{\Psi}=\partial_{t}\Psi.
\end{equation}
If we write $\Psi(t)=\psi(\boldsymbol{\theta})$ , with $\boldsymbol{\theta}=\boldsymbol{\omega}t$,
the above equation can be rewritten as
\begin{equation}
i\boldsymbol{\omega}\cdot\boldsymbol{\nabla}_{\boldsymbol{\theta}}\psi(\boldsymbol{\theta})=h(\boldsymbol{\theta})\psi(\boldsymbol{\theta}).
\end{equation}

Let us formally divide $\boldsymbol{\theta}$ into two parts as
$\boldsymbol{\theta}=(\boldsymbol{\theta}_{\perp},\theta_{k})$, 
where $\boldsymbol{\theta}_{\perp}$ is a vector consisting of $\theta_{j}$s with $j=1,\dots,d, j\neq k$.
Similarly, we write $\boldsymbol{\omega}=(\boldsymbol{\omega}_{\perp},\omega_k)$. 

Thus, we obtain a new SEQ
\begin{equation}
i\omega_k\partial_{\theta_k}\psi(\boldsymbol{\theta}_{\perp},\theta_k )=\left[h(\boldsymbol{\theta}_{\perp},\theta_{k}
)-i\boldsymbol{\omega}_{\perp}\cdot\boldsymbol{\nabla}_{\theta_{\perp}}\right]\psi(\boldsymbol{\theta}_{\perp},\theta_{k}).
\end{equation}
By Floquet theorem, the solutions to this SEQ can be written as
\begin{equation}
\psi(\boldsymbol{\theta}_{\perp},\theta_k)=\exp(-i\epsilon_k \theta_k/\omega_k)\phi_{d}(\boldsymbol{\theta}_{\perp},\theta_k)
\end{equation}
with $\phi_{k}(\boldsymbol{\theta}_{\perp},\theta_{k})=\phi_{k}(\boldsymbol{\theta}_{\perp},\theta_{k}+2\pi)$.
Hence, $\psi(\boldsymbol{\theta})\exp(i\epsilon_k\theta_k/\omega_k)$ is $2\pi$-periodic in its $k$th argument $\theta_k$. 
Since $k$ is an arbitrary number from $1$ to $d$, 
\begin{equation}
\phi(\boldsymbol{\theta}) = \psi(\boldsymbol{\theta}) \exp(i\sum_{j=1}^d \epsilon_j \theta_j/\omega_j)
\end{equation}
will be $2\pi$-periodic in all $\theta_j$s with proper chosen $\epsilon_j$s. 

As a result, a quasiperiodic function $\Phi(t)=\phi(\boldsymbol{\theta})$ can be constructed by setting
$\boldsymbol{\theta}=\boldsymbol{\omega}t$. We thus obtain a factorization 
\begin{equation}
\Psi(t) = \Phi(t)\exp(-i\epsilon t), \quad \epsilon = \sum_{j=1}^d \epsilon_j, 
\end{equation}
with $\Phi(t)$ time-quasiperiodic in the same frequencies.  Moreover, this function satisfies
\begin{equation}
\left[H(t) - i\partial_t\right]\Phi(t) = \epsilon \Phi(t),
\end{equation}
which is Eq.~(6) in the main text.

\section*{Wannier-Stark localization of Floquet Majoranas}
Let us consider the time-periodic Kitaev chain introduced in the main text, with Hamiltonian $H(t) = H_{K} + M(\omega
t)$. The static part is
\begin{equation}
  H_{K} = -\mu\sum_{j=1}^N c_{j}^\dagger c_{j} -\sum_{j=1}^{N-1}[(Jc_{j}^\dagger c_{j+1} + i\Delta c_{j} c_{j+1}) +
  h.c.],
  \label{seq:kitaev}
\end{equation}
and the time-periodic part is
\begin{equation}
  M(\omega t) = -i\Delta' \sum_{j=1}^{N-1}( e^{-i\omega t} c_{j} c_{j+1} - e^{i\omega t}c_{j+1}^\dagger c_{j}^\dagger).
  \label{seq:M_pairing}
\end{equation}

Introducing Nambu spinor $C_j^\dagger = (c_j^\dagger, c_j)$, we obtain the corresponding Bogoliubov--de Gennes
Hamiltonian up to a constant term
\begin{gather}
H_{BdG}  = H_{BdG,K} + M_{BdG}(\omega t) \\
H_{BdG,K} = -\frac{\mu}{2}\sum_{j=1}^{N}C_{j}^\dagger\tau_z C_j - \frac{1}{2}\sum_{j=1}^{N-1}\left[ 
C_{j}^\dagger\left(J\tau_z +i\Delta\tau_x\right)C_{j+1} + h.c.\right] \\
M_{BdG}(\omega t) = -\frac{i\Delta'}{2} \sum_{j=1}^{N-1} \left[ C_{j}^{\dagger} e^{i\omega t\tau_z}\tau_x C_{j+1} + h.c.
\right].
\end{gather}
If we rather consider a periodic boundary condition and take the Fourier expansion $C_j = \sum_{j=1}^N \Psi(k) e^{i k
j}/\sqrt{N}$, we obtain the Bloch Hamiltonian given the in main text.

This time-periodic Hamiltonian can be mapped to a 1D synthetic lattice with an additional electric field,
giving rise to a Wannier-Stark ladder. The on-site Hamiltonian at the $n$th rung is $h_{0}-n\omega\mathbb{I}_{2N}$, with
\begin{equation}
h_0 = -\frac{1}{2}\left(\begin{array}{ccccc}
\mu\tau_{z} & J\tau_{z}+i\Delta\tau_{x}\\
J\tau_{z}-i\Delta\tau_{x} & \mu\tau_{z} & J\tau_{z}+i\Delta\tau_{x}\\
 & J\tau_{z}-i\Delta\tau_{x} & \ddots & \ddots\\
 &  & \ddots & \mu\tau_{z} & J\tau_{z}+i\Delta\tau_{x}\\
 &  &  & J\tau_{z}-i\Delta\tau_{x} & \mu\tau_{z}
\end{array}\right)
\end{equation}
a $2N\times2N$ matrix describing a finite Kitaev chain of length
$N$ (in unit of lattice constant), and $\mathbb{I}_{2N}$ is the identity matrix of the same size. 
The nearest-neighbor hopping matrix along the ladder (from the $n$th to the $(n+1)$th rung of the ladder) is
\begin{equation}
h_{-1} = -\frac{i\Delta'}{2}\left(\begin{array}{ccccc}
0 & \tau_{+}\\
-\tau_{+} & 0 & \tau_{+}\\
 & -\tau_{+} & \ddots & \ddots\\
 &  & \ddots & 0 & \tau_{+}\\
 &  &  & -\tau_{+} & 0
\end{array}\right),
\end{equation}
with $\tau_{\pm} = (\tau_{x} \pm i\tau_y)/2$.
Hopping in the opposite direction is given by the matrix $h_1 = h_{-1}^\dagger$.  

\begin{figure}[t]
\centering
\includegraphics[width=0.45\textwidth]{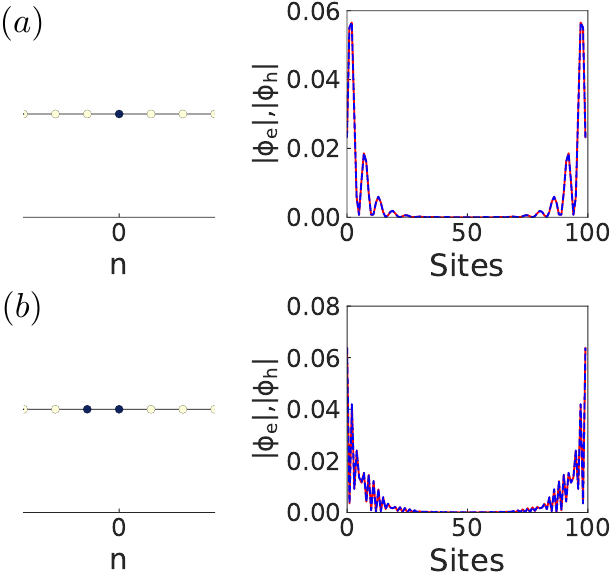}
\caption{\label{sfig:F-majorana} Numerical results for the Floquet Majorana wave functions in a Wannier-Stark ladder of
$30$ rungs, for a time-periodic Kitaev chain of $N=100$ sites. The left panels are the magnitude of $\abs{\phi_{n}})$
(summed over electron and hole components), where darker color corresponds to larger magnitude. The right panels are
the absolute value of the corresponding Majorana wave function, summed over the 1D synthetic lattice. The electron
and hole components $\phi_e$ and $\phi_h$ are plotted as red solid and blue dashed curves. $(a)$ and $(b)$
are for Majoranas at quasienergies $0$ and $\omega/2$, respectively.
The other parameters are $J/\omega=0.51$, $\mu/\omega=0.87$, and $\Delta/\omega=0.051$.
}
\end{figure}

In Fig.~\ref{sfig:F-majorana}, we numerically calculate the Floquet Majorana wave function
$\phi_{n}(j)=(\phi_{n,e}(j), \phi_{n,h}(j))$ at quasienergies $0$ and $\omega/2$
in a Kitaev chain of $N=100$ sites. We take $30$ rungs of the Wannier-Stark ladder in our numerical simulation.
We see that both Majoranas are perfectly localized in both physical space and the synthetic lattice.

\section*{Localization in a quasiperiodic ladder}
When the $d$-time-quasiperiodic system is mapped to a $d$ dimensional synthetic lattice, the presence of the electric field
$\boldsymbol{\omega}$ naturally cuts the lattice into a layers of quasicrystals living in one dimension lower. 
These quasicrystals are constructed by taking all the lattice points in between two equipotential surfaces perpendicular
to the electric field, as described in the main text. Hence, the on-site potentials of the quasicrystal
stays close to the average potential of the two surfaces. On the other hand, 
the on-site potential within the quasicrystal varies from site to site. For two neighboring sites, the 
potential difference is one of $\omega_j$s for $j=1,\dots d$.
Thus, this quasiperiodic structure can be viewed as a mixuture of $d$ Wannier-Stark ladders, which
stays at a constant height in average. When the potential difference 
is larger compared to the coupling strength between neighboring rungs in this mixed ladder, the eigenstates of
the system are localized. 

\begin{figure}[t]
\centering
\includegraphics[width=0.45\textwidth]{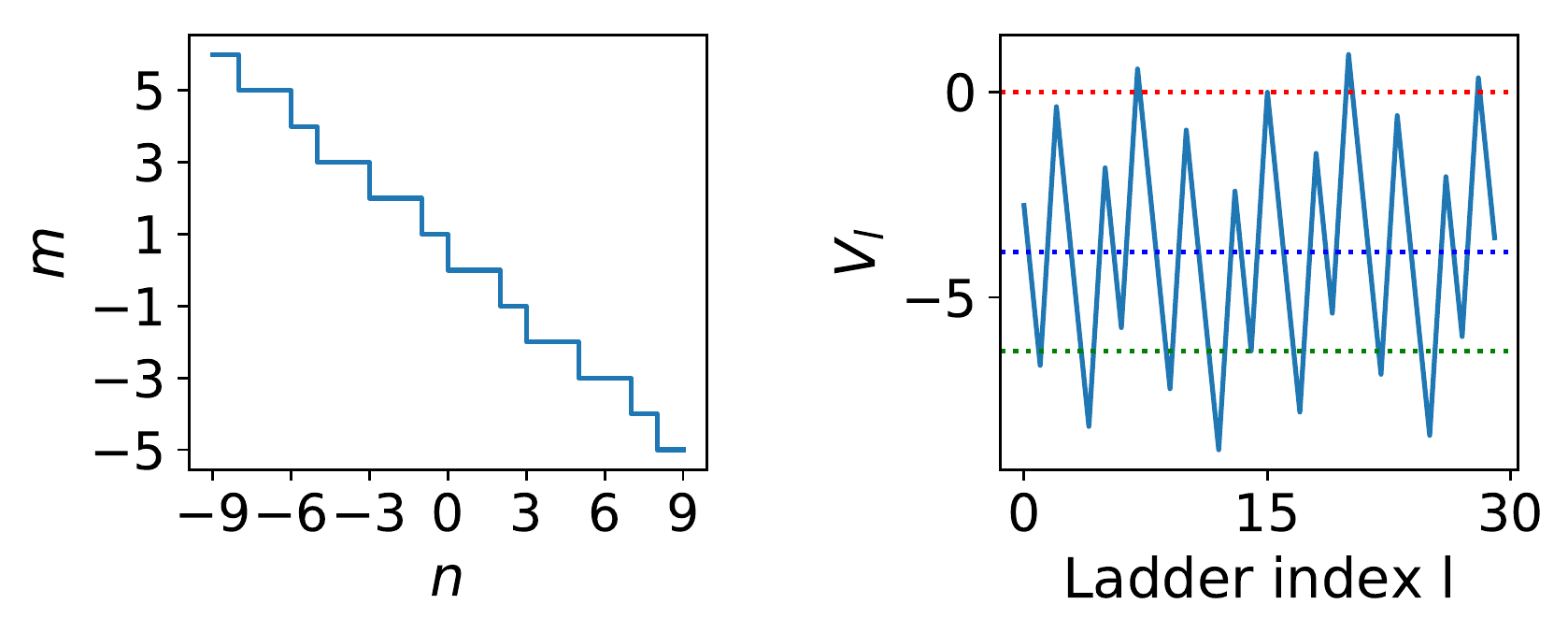}
\caption{\label{sfig:quasicrystal-chain} Left: 1D quasiperiodic ladder obtained by cutting the 2D synthetic lattice with
equipotential surfaces. Right: Onsite potential $V_l$ as a function of the ladder index $l$. 
We indicate the energy at $0$, $-\omega_1$, $-\omega_2$ by the red, blue and green dotted lines for reference.  }
\end{figure}

The time-quasiperiodic Kitaev chain introduced in the main text can be mapped to a 2D synthetic lattice with 
an additional electric field. Perpendicular to the field, we have a quasiperiodic ladder
climbing up or down by either $\omega_1$ or $\omega_2$ between two rungs, depending on 
whether these two rungs are connected horizontally or vertically in the original 2D synthetic lattice.
In Fig.~\ref{sfig:quasicrystal-chain} we show a quasiperiodic ladder of length $30$ obtained in a 2D lattice,
and its on-site potential $V_l$ as a function the ladder index $l$. 

Moreover, the hopping matrix (from the $n$th to  the $(n+1)$th rung of the ladder) is 
\begin{equation}
h_{-1,0} = -\frac{i\Delta'}{2}\left(\begin{array}{ccccc}
0 & \tau_{+}\\
-\tau_{+} & 0 & \tau_{+}\\
 & -\tau_{+} & \ddots & \ddots\\
 &  & \ddots & 0 & \tau_{+}\\
 &  &  & -\tau_{+} & 0
\end{array}\right)
\end{equation}
for a horizontal hopping, or 
\begin{equation}
h_{0,1} = -\frac{i\Delta'}{2}\left(\begin{array}{ccccc}
0 & \tau_{-}\\
-\tau_{-} & 0 & \tau_{-}\\
 & -\tau_{-} & \ddots & \ddots\\
 &  & \ddots & 0 & \tau_{-}\\
 &  &  & -\tau_{-} & 0
\end{array}\right)
\end{equation}
for a vertical hopping. 

\begin{figure}[t]
\centering
\includegraphics[width=0.7\textwidth]{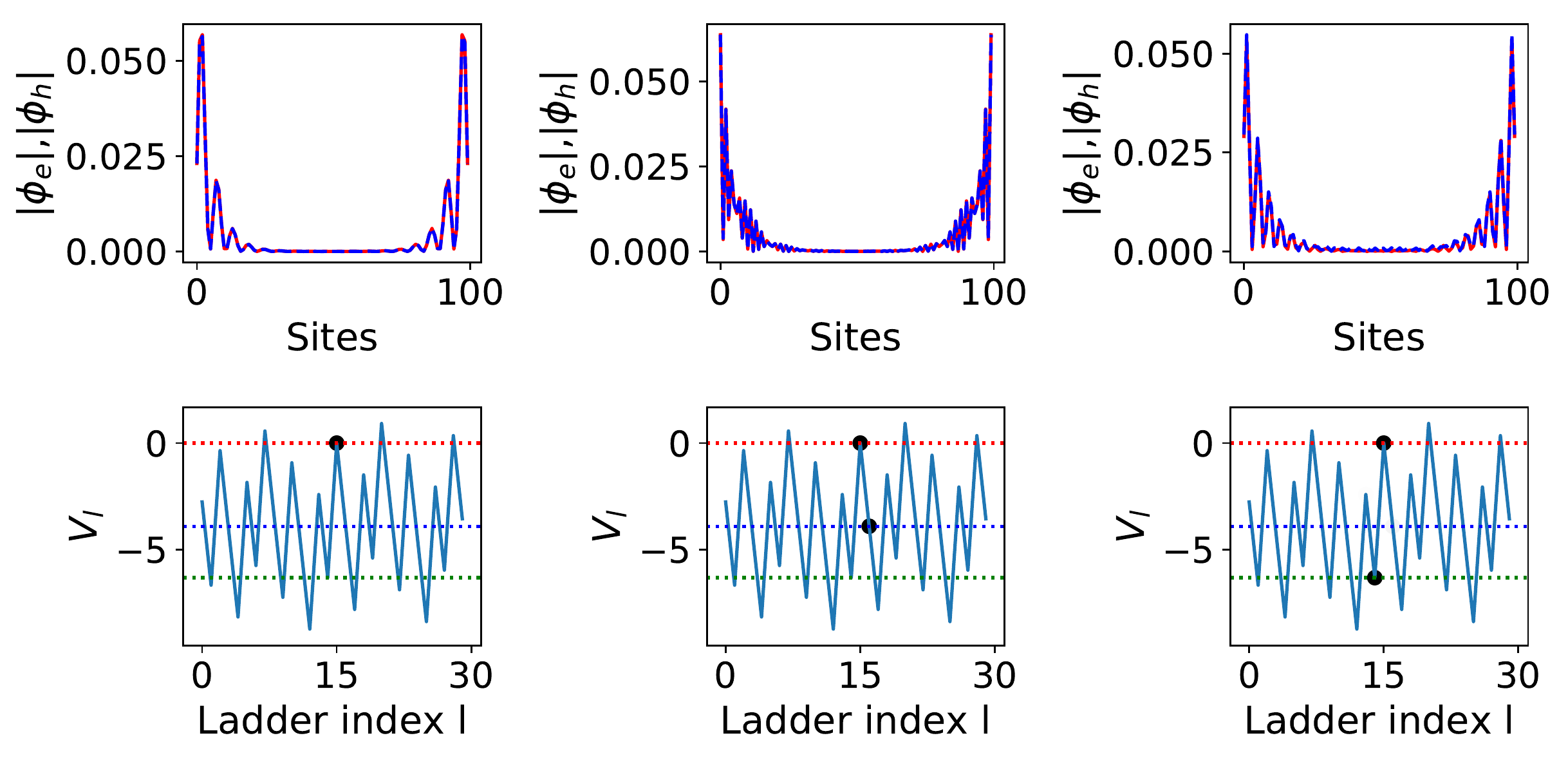}
\caption{\label{sfig:quasicrystal-Majorana} Majorana wave functions in a quasiperiodic ladder of 30 rungs, where each
rung contains a Kitaev chain of $N=100$ sites. Upper: the absolute value of the Majorana wave function, summed over the
quasiperiodic ladder. The electron and hole components $\phi_e$ and $\phi_h$ are plotted as red solid and blue dashed curves.
Lower: the magnitude of $\abs{\phi_{j}})$ (summed over electron and hole components) plotted 
on top of the on-site potential as a function of the ladder index, where darker color corresponds to larger magnitude. 
The left, middle, and right panels are for Majoranas at $0$, $-\omega_1/2$ and $-\omega_2/2$ energies, respectively.
The other parameters are $\omega_1=3.9$, $\omega_2 = \omega_1\times(\sqrt{5}+1)/2$, $J=2$, $\mu=3.4$, $\Delta=0.2$, and
$\Delta'=0.15$. 
}
\end{figure}

In Fig.~\ref{sfig:quasicrystal-Majorana}, we numerically calculate the Majorana wave function
$\phi_{l}(j)=(\phi_{l,e}(j), \phi_{l,h}(j))$ ($l$ is the ladder index) at quasienergies $0$, $-\omega_1/2$, and $-\omega_2/2$
in a Kitaev chain of $N=100$ sites. We take $30$ rungs of the quasiperiodic ladder in our numerical simulation.
We see that both Majoranas are perfectly localized in both physical space and the quasiperiodic ladder.

Indeed, combining the two localization mechanisms, i.e., the Wannier-Stark localization and quasiperiodic localization, 
time-quasiperiodic Majoranas can be localized in the synthetic dimensions as discussed in the main text. 

\section*{Particle-hole symmetry of time-quasiperiodic Majoranas}
In this section, we numerically show that the time-quasiperiodic Majorana wave functions are particle-hole symmetric at all times. 

The Majorana wave function $\psi_{j}(x,t)$ at position $x$ and time $t$ can be written as
\begin{equation}
\psi_j (x,t) = e^{-i\epsilon_j t} \sum_{n_1,n_2}e^{-i(n_1\omega_1+n_2\omega_2)t} \phi_j(x;n_1,n_2), \quad j=0,1,2,
\end{equation}
with $\epsilon_0 = 0$ and $\epsilon_j = \omega_j/2$ for $j=1,2$. Here $\psi_j(x,t)$ is a two-component wave function
consisting electron and hole components $\psi_{j}^{e}(x,t)$ and $\psi_{j}^h(x,t)$.  To show the particle-symmetry of
$\psi_j(x,t)$ at any time $t$, one can compute the difference
\begin{equation}
\chi(x,t) = \abs{\psi_{j}^e(x,t)}^2 - \abs{\psi_{j}^h(h,t)}^2
\end{equation}
and show it vanishes at all $x$ and $t$.

\begin{figure}[h]
\centering
\includegraphics[width=0.8\textwidth]{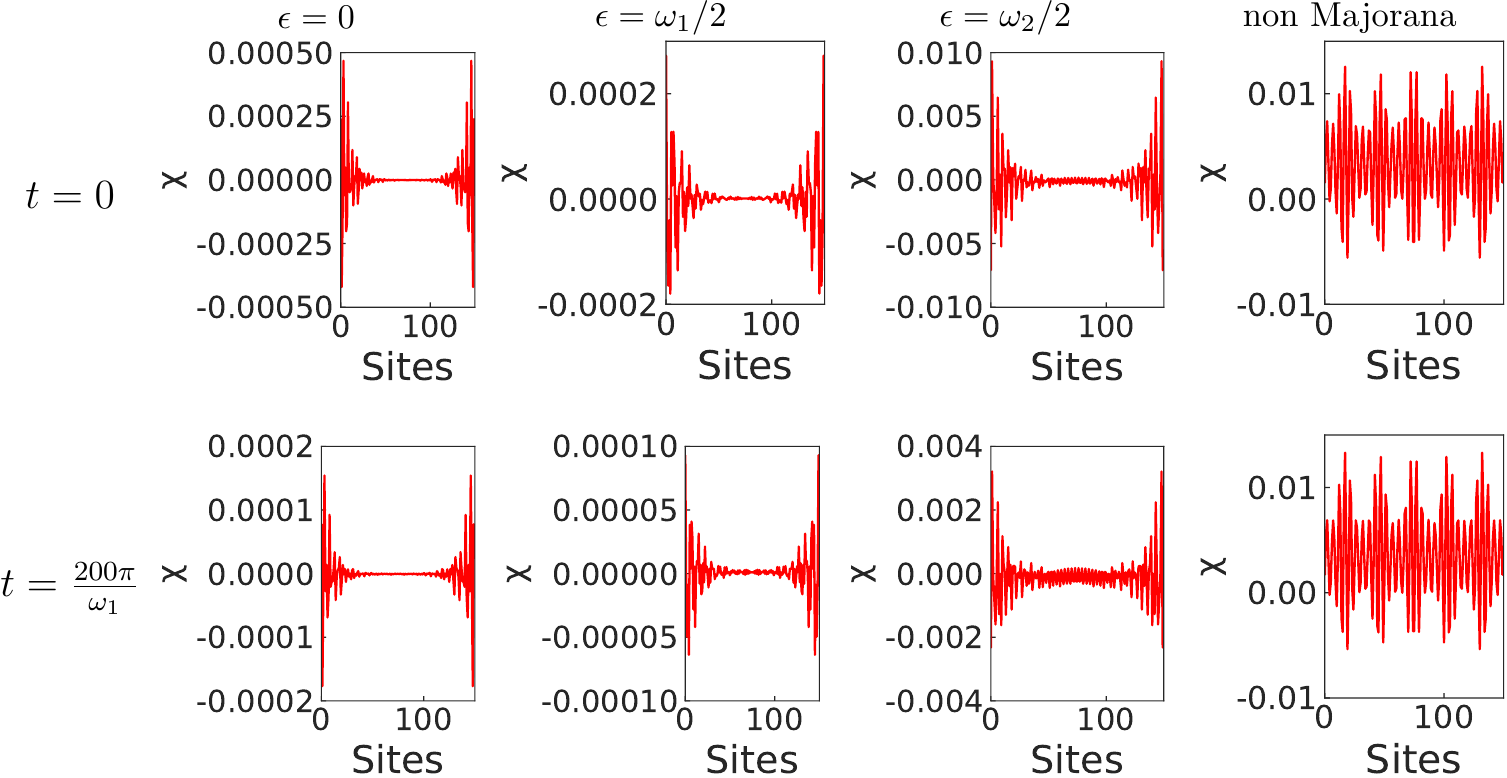}
\caption{Difference in electron and hole components $\chi$ of the time-quasiperiodic Majoranas with energy $\epsilon$
(first three columns) as well as a generic non Majorana state (last column), at times $t=0$ and $t = 100 \times 2\pi/\omega_1$ (two rows).
The parameters are: $\omega_2/\omega_1= (\sqrt{5}+1)/2$, $J/\omega_1 = 0.51$, $\mu/\omega_1=0.87$,
$\Delta/\omega_1=0.051$ and $\Delta'/\omega_1 = 0.038$ for a chain of $L=150$ sites.  }
\label{sfig:TDM}
\end{figure}

\begin{figure}[h]
\centering
\includegraphics[width=0.7\textwidth]{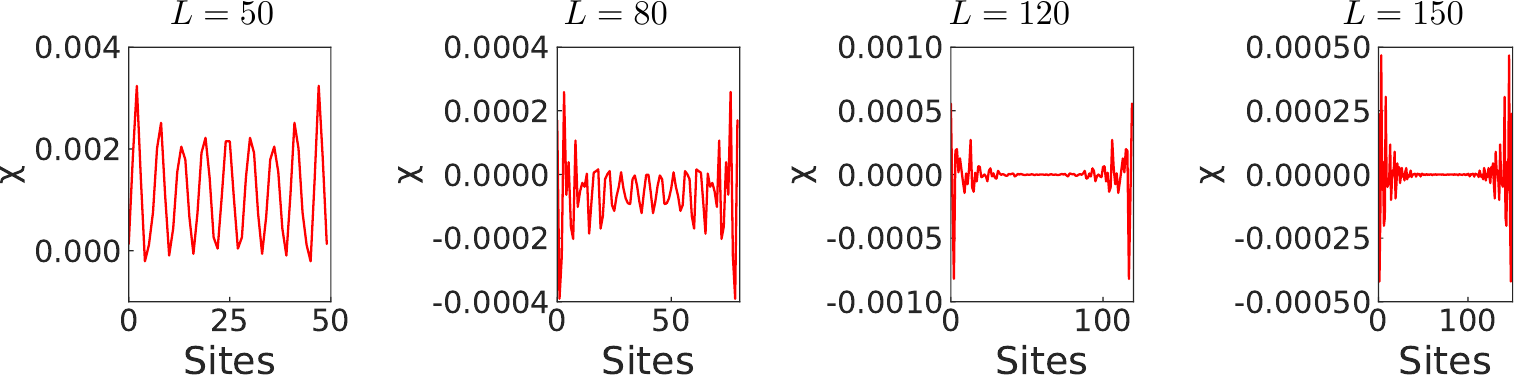}
\caption{Difference in electron compoenent and hole component $\chi$ of the time-quasiperiodic Majoranas with energy
$\epsilon = 0$ at time $t=0$, computed with different chain lengths $L$. The parameters are the same as the ones in
Fig.~\ref{sfig:TDM}. }
\label{sfig:finitesize}
\end{figure}

In Fig.~\ref{sfig:TDM}, we show the difference in electron and hole components $\chi$ of the time-quasiperiodic Majoranas with energy $\epsilon$
in the first three columns, at times $t=0$ and $t=100\times 2\pi/\omega_1$, in two rows. 
We see that $\chi$ for time-quasiperiodic Majoranas is very close to zero, compared to $\chi$ of a generic non Majorana
state. In fact, the small deviation of $\chi$ from zero is due to finite size effect of the 1D chain we used in our
numerical calculation. In Fig.~\ref{sfig:finitesize}, we show different $\chi$s computed with different chain lengths. 
We see that indeed as the number of sites $L$ along the chain increases, $\chi$ approches zero.

\section*{More general discussion with time-dependent chemical potential and hopping}

\subsection*{Floquet system and Wannier-Stark localization}

Time dependent chemical potential and hopping term can be characterized
by the time-dependent function $\xi_{k}(t)=\epsilon_{k}(t)-\mu(t)$,
where $k$ is the momentum when considering periodic boundary condition
along the 1D superconductor, $\epsilon_{k}(t)$ is the time-dependent
band structure, and $\mu(t)$ is the time-dependent chemical potential.
For simplicity, let us consider 
\begin{equation}
\xi_{k}(t)=\xi_{k}+2J'\cos\omega t.
\end{equation}
The time-periodic Kitaev chain in general can be written as
\begin{equation}
H_{k}(t)=\xi_{k}(t)\tau_{z}+\Delta_{k}(t)\tau_{x},
\end{equation}
with $\Delta_{k}(t)$ a time-periodic pairing function. For each $k$,
the time-dependent Hamiltonian can be mapped to a Wannier-Stark ladder
of two level systems according to Eq. (1) in the main text. One can
first neglect the superconducting pairing potential $\Delta_{k}(t)\tau_{x}$,
and focus on the normal dispersion only. The Schr\"odinger equation
corresponding to the mapped time-independent system can be written
as
\begin{equation}
J'\tau_{z}(\phi(n+1)+\phi(n-1))+(\xi_{k}\tau_{z}-n\omega)\phi(n)=E\phi(n),
\end{equation}
where $\phi(n)$ is the two component wave function amplitude at the
$n$th rung at energy $E$. 

If we define dimensionless quantity $\alpha=J'/\omega$, $\tilde{\xi_{k}}=\xi_{k}/\omega$,
$\epsilon=E/\omega$, the above equation can be rewritten as
\begin{equation}
\phi(n+1)+\phi(n-1)=\frac{\epsilon\tau_{z}-\tilde{\xi_{k}}+n\tau_{z}}{\alpha}\phi(n).
\end{equation}
Recall the recurrence relation for Bessel function
\begin{equation}
Z_{n+1}(x)+Z_{n-1}(x)=\frac{2n}{x}Z_{n}(x)
\end{equation}
where $Z_{n}(x)$ can be the Bessel function of the first kind $J_{n}(x)$
or of the second kind $N_{n}(x)$. Hence, we require
\begin{equation}
\pm\epsilon-\tilde{\xi}_{k}\in\mathbb{Z}.
\end{equation}
We can label these energies as
\begin{equation}
\epsilon_{m}^{+}=m+\tilde{\xi}_{k},\quad\epsilon_{l}^{-}=l-\tilde{\xi}_{k},
\end{equation}
with $l,m\in\mathbb{Z}$. 
\begin{equation}
\phi_{n+1}+\phi_{n-1}=\frac{-l-n}{\alpha}\phi_{n}
\end{equation}
\begin{equation}
\end{equation}
Since the wave function needs to be normalizable, we get two set of
solutions
\begin{equation}
\phi_{m}^{+}(n)\equiv\braket{n|\phi_{m}^{+}}=\left(\begin{array}{c}
J_{m+n}(2\alpha)\\
0
\end{array}\right)\quad{\rm and\quad}\phi_{l}^{-}(n)\equiv\braket{n|\phi_{l}^{-}}=\left(\begin{array}{c}
0\\
J_{-l-n}(2\alpha)
\end{array}\right),
\end{equation}
 corresponding states at energies $\epsilon_{m}^{+}$ and $\epsilon_{l}^{-}$.
Given the fact that for small arguement $0<z\ll\sqrt{\alpha+1}$
\begin{equation}
J_{\alpha}(z)\simeq\frac{1}{\alpha!}\left(\frac{z}{2}\right)^{\alpha},\label{eq:Bessel-asymptotic}
\end{equation}
we see that $\phi_{m}^{\pm}$ are localized at the $-m$th rung of
the Wannier-Stark ladder (Wannier-Stark localization). 

Let us now take into account the time-periodic pairing potential $\Delta_{k}(t)\tau_{x}$,
which creates coupling between states $\ket{\phi_{m}^{\pm}}$in the
mapped time-independent problem. Assuming
\begin{equation}
\Delta_{k}(t)=\sum_{n\in\mathbb{Z}}e^{-in\omega t}\Delta_{k}^{(n)},
\end{equation}
then the only nonzero matrix elements are
\begin{align}
\bra{\phi_{m}^{+}}\hat{\Delta}_{k}\tau_{x}\ket{\phi_{l}^{-}} & =\sum_{nn'\in\mathbb{Z}}\braket{\phi_{m}^{+}|n}\bra{n}\hat{\Delta}_{k}\tau_{x}\ket{n'}\braket{n'|\phi_{l}^{-}}\nonumber \\
 & =\sum_{nn'\in\mathbb{Z}}J_{m+n}(2\alpha)J_{-l-n'}(2\alpha)\Delta_{k}^{(n-n')}\nonumber \\
 & =\sum_{s\in\mathbb{Z}}\Delta_{k}^{(s)}\sum_{n\in\mathbb{Z}}J_{m+n}(2\alpha)J_{-l-n+s}(2\alpha)\nonumber \\
 & =\sum_{s\in\mathbb{Z}}\Delta_{k}^{(s)}\sum_{r\mathbb{\in Z}}J_{r}(2\alpha)J_{m-l+s-r}(2\alpha)\nonumber \\
 & =\sum_{s\in\mathbb{Z}}\Delta_{k}^{(s)}J_{m-l+s}(4\alpha)=\bra{\phi_{m-l}^{+}}\hat{\Delta}_{k}\tau_{x}\ket{\phi_{0}^{-}}\equiv D_{m-l},
\end{align}
where we used the Bessel function addition theorem
\begin{equation}
\sum_{m\in\mathbb{Z}}J_{n-m}(x)J_{m}(y)=J_{n}(x+y).
\end{equation}

To create Majoranas at zero quasienergy, we need $\epsilon_{0}^{\pm}$
cross at some $k$, at which $D_{0}\neq0$. Similarly, to have Majoranas
at $\omega/2$ quasienergy, we require, for example $\epsilon_{0}^{+}$
crosses $\epsilon_{1}^{-}$ at some $k$ when $D_{-1}\neq0$. Even
if we take static pairing $\Delta_{k}(t)=\Delta\sin k$, namely $\Delta_{k}^{(s)}=\delta_{s0}\Delta_{k}$
and $D_{r}=J_{r}(4\alpha)$, we can have both types of Majoranas,
for example taking $J_{0}(4\alpha)\simeq J_{1}(4\alpha)\gg J_{\nu}(4\alpha)$
with $\nu\geq2\in\mathbb{Z}$. 

\subsection*{Localization in time-quasiperiodic system}

We now generalization the previous Floquet superconductor to a two-frequency-time-quasiperiodic
superconductor. By using two-dimensional Bessel functions \cite{Korsch2006} and performing
similar analysis, we will show localization in the mapped two dimensional
time-independent synthetic lattice. We will then construct time-quasiperiodic
topological superconductor with Majorana multiplexing.

Consider time-dependent dispersion as a function of time $t$ and
momentum $k$
\begin{equation}
\xi_{k}(t)=\xi_{k}+2J'_{1}\cos\omega_{1}t+2J_{2}'\cos\omega_{2}t
\end{equation}
for simplicity. The time-quasiperiodic Kitaev chain we consider is
\begin{equation}
H_{k}(t)=\xi_{k}(t)\tau_{z}+\Delta_{k}(t)\tau_{x},
\end{equation}
where $\Delta_{k}(t)$ is a time-quasiperiodic function at frequencies
$\omega_{1}$ and $\omega_{2}$. It is helpful to first consider rational
case with $\omega_{1}/\omega_{2}=p/q$ with coprime intergers $p$
and $q$. The time-quasiperiodic case with irrational $\omega_{1}/\omega_{2}$
can be regarded as the limiting procedure
\begin{equation}
\frac{\omega_{1}}{\omega_{2}}=\lim_{p_{n},q_{n}\to\infty}\frac{p_{n}}{q_{n}}.
\end{equation}

For each $k$, the two-frequency-time-quasiperiodic system can be
mapped to a two-dimensional synthetic lattice, in which each lattice
site corresponding to a two level system. Let us denote the two component
wave function amplitude at the site $(n,m)$ as $\phi(n,m)$, the
Schr\"odinger equation can then be written as
\begin{equation}
J_{1}'\tau_{z}(\phi(n+1,m)+\phi(n-1,m))+J_{2}'\tau_{z}(\phi(n,m+1)+\phi(n,m-1))+(\xi_{k}\tau_{z}-n\omega_{1}-m\omega_{2})\phi(n,m)=E\phi(n,m).
\end{equation}
We further more introduce dimensionless quantities
\begin{equation}
\alpha_{1}=\frac{J_{1}'}{\omega_{1}},\quad\alpha_{2}=\frac{J_{2}'}{\omega_{2}},\quad\epsilon=\frac{E}{\omega_{1}}p=\frac{E}{\omega_{2}}q,\quad\tilde{\xi}_{k}=\frac{\xi_{k}}{\omega_{1}}p=\frac{\xi_{k}}{\omega_{2}}q.
\end{equation}
We can rewrite the above equation as
\begin{equation}
p\alpha_{1}\tau_{z}(\phi(n+1,m)+\phi(n-1,m))+q\alpha_{2}\tau_{z}(\phi(n,m+1)+\phi(n,m-1))=(\epsilon-\tilde{\xi}_{k}\tau_{z}+pn+qm)\phi(n,m).\label{eq:time-xi-seq}
\end{equation}

We introduce the two dimensional Bessel function \cite{Korsch2006}
\begin{equation}
J_{n}^{p,q}(u,v)=\frac{1}{2\pi}\int_{-\pi}^{\pi}dt\,e^{i(u\sin pt+v\sin qt-nt)}=\frac{1}{\pi}\int_{0}^{\pi}dt\,\cos\left(u\sin pt+v\sin qt-nt\right),
\end{equation}
which fulfill the following recurrence relation
\begin{equation}
pu(J_{\nu-p}^{p,q}(u,v)+J_{\nu+p}^{p,q}(u,v))+qv(J_{\nu-q}^{p,q}(u,v)+J_{\nu+q}^{p,q}(u,v))=2\nu J_{\nu}^{p,q}(u,v).\label{eq:2dbessel}
\end{equation}
Compairing Eq.(\ref{eq:time-xi-seq}) with (\ref{eq:2dbessel}), we
finds two sets of solutions
\begin{equation}
\phi_{r}^{+}(n,m)=\left(\begin{array}{c}
J_{r+pn+qm}^{p,q}(2\alpha_{1},2\alpha_{2})\\
0
\end{array}\right),\quad\phi_{s}^{-}(n,m)=\left(\begin{array}{c}
0\\
J_{-s-pn-qm}^{p,q}(2\alpha_{1},2\alpha_{2})
\end{array}\right),\quad r,s\in\mathbb{Z},\label{eq:2dwf}
\end{equation}
with eigenvalues
\begin{equation}
\epsilon_{r}^{+}=r+\tilde{\xi}_{k},\quad\epsilon_{s}^{-}=s-\tilde{\xi}_{k}.
\end{equation}

Let us take a closer look into the two dimensional Bessel function
can be represented in terms of ordinary Bessel function \cite{Korsch2006}
\begin{equation}
J_{\nu}^{p,q}(u,v)=\sum_{(N,M)\in S_{\nu}}J_{N}(u)J_{M}(v),
\end{equation}
where the sum is over all pairs $(N,M)$ in the set of solutions 
\begin{equation}
S_{\nu}=\{(N,M)|pN+qM=\nu\}
\end{equation}
 of the Diophantine equation 
\begin{equation}
pN+qM=\nu.
\end{equation}
If $(N_{0},M_{0})$ is a particular solution of the above equation,
which can be found by the Euclidean algorithm, then all solutions
can be written as $(N_{0}-qw,M_{0}+pw)$ with $w\in\mathbb{Z}$. By
Eq.(\ref{eq:Bessel-asymptotic}), we see that $J_{\nu}^{pq}(u,v)$
is mainly contributed from $\abs{N}\lesssim u$, $\abs{M}\apprle v$.
In particular, when $q\gg u,$ $p\gg v$ as we will consider the irrational
limit, there is at most one solution $(N_{0},M_{0})$ satisfying $\abs{N_{0}}\lesssim u,\abs{M_{0}}\apprle v$,
since $\abs{N_{0}-qw}\gg u$, $\abs{M_{0}+pw}\gg v$ for $w\neq0$.
When such a solution exist, we have $J_{\nu}^{p,q}(u,v)\simeq J_{N_{0}}(u)J_{M_{0}}(v)$;
otherwise $J_{\nu}^{p,q}(u,v)$ is very small. In other words, as
we change $\nu$, like the ordinary Bessel function, $J_{\nu}^{p,q}(u,v)$
is localized around $\nu=pN_{0}+qM_{0}$ with $(N_{0},M_{0})\in S_{\nu}$
with $\abs{N_{0}}\lesssim u,\abs{M_{0}}\apprle v$. 

By Eq.(\ref{eq:2dwf}), we see that the wave function amplitude are
the same at sites $(n,m)$ with constant $pn+qm$, which are sites
along the direction $(-q,p)$ perpendicular to the field direction
$(p,q)$. Combining the properties of the two dimensional Bessel function
discussed above, we know that $\phi_{r}^{\pm}(n,m)$ are actually
localized around $(n,m)$ when there exists
\begin{equation}
(N_{0},M_{0})\in S_{r}+(n,m)
\end{equation}
with $\abs{N_{0}}\lesssim2\alpha_{1},\abs{M_{0}}\apprle2\alpha_{2}$.
Since the separation between peaks in the wave function amplitudes
is $(-q,p)$ in the 2D synthetic lattice, in the irrational limit,
we actually have true quasiperiodic localization along $(-q,p)$.
Along the direction of the field $(p,q)$, the states are also localized,
which is understood as the Wannier-Stark localization.

Let us now take into account the time-periodic pairing potential $\Delta_{k}(t)\tau_{x}$,
which creates coupling between states $\ket{\phi_{r}^{\pm}}$in the
mapped time-independent problem. Assuming
\begin{equation}
\Delta_{k}(t)=\sum_{n_{1}\in\mathbb{Z}}e^{-i(n_{1}\omega_{1}+n_{2}\omega_{2})t}\Delta_{k}^{(n_{1},n_{2})},
\end{equation}
then the only nonzero matrix elements are
\begin{align}
\bra{\phi_{r}^{+}}\hat{\Delta}_{k}\tau_{x}\ket{\phi_{s}^{-}} & =\sum_{n_{1},n_{2},n_{1}',n_{2}'\in\mathbb{Z}}\braket{\phi_{r}^{+}|n_{1}n_{2}}\bra{n_{1}n_{2}}\hat{\Delta}_{k}\tau_{x}\ket{n_{1}'n_{2}'}\braket{n_{1}n_{2}|\phi_{s}^{-}}\nonumber \\
 & =\sum_{n_{1},n_{2},n_{1}',n_{2}'\in\mathbb{Z}}J_{r+pn_{1}+qn_{2}}^{p,q}(2\alpha_{1},2\alpha_{2})J_{-s-pn_{1}'-qn_{2}'}^{p,q}(2\alpha_{1},2\alpha_{2})\Delta_{k}^{(n_{1}-n'_{1},n_{2}-n_{2}')}\nonumber \\
 & =\sum_{m_{1},m_{2}\in\mathbb{Z}}\Delta_{k}^{(m_{1},m_{2})}\sum_{n_{1},n_{2}\in\mathbb{Z}}J_{r+pn_{1}+qn_{2}}^{p,q}(2\alpha_{1},2\alpha_{2})J_{-s+pm_{1}+qm_{2}-pn_{1}-qn_{2}}^{p,q}(2\alpha_{1},2\alpha_{2})\nonumber \\
 & =\sum_{m_{1},m_{2}\in\mathbb{Z}}\Delta_{k}^{(m_{1},m_{2})}\sum_{\nu\in\mathbb{Z}}J_{r+\nu}^{p,q}(2\alpha_{1},2\alpha_{2})J_{-s+pm_{1}+qm_{2}-\nu}^{p,q}(2\alpha_{1},2\alpha_{2})\nonumber \\
 & =\sum_{m_{1},m_{2}\in\mathbb{Z}}\Delta_{k}^{(m_{1},m_{2})}J_{pm_{1}+qm_{2}+r-s}^{p,q}(4\alpha_{1},4\alpha_{2})\nonumber \\
 & =\bra{\phi_{r-s}^{+}}\hat{\Delta}_{k}\tau_{x}\ket{\phi_{0}^{-}}\equiv F_{r-s},
\end{align}
where we used the B\'ezout's identity \cite{Jones1998}
\begin{equation}
\{pn_{1}+qn_{2}|n_{1},n_{2}\in\mathbb{Z},\gcd(p,q)=1\}=\mathbb{Z}
\end{equation}
and the addition theorem for the two dimensional Bessel function \cite{Korsch2006}
\begin{equation}
\sum_{m\in\mathbb{Z}}J_{n-m}^{p,q}(u_{1},v_{1})J_{m}^{p,q}(u_{2},v_{2})=J_{n}^{p,q}(u_{1}+u_{2},v_{1}+v_{2}).
\end{equation}
Note that if $\phi_{r}^{\pm}(n,m)$ is localized around $(n_{0},m_{0})$,
then $\phi_{r-p}^{\pm}(n,m)$ is localized around $(n_{0}+1,m_{0})$,
and $\phi_{r-q}^{\pm}(n,m)$ is localized around $(n_{0},m_{0}+1)$.

To create Majoranas at zero quasienergy, we need $\epsilon_{0}^{\pm}$
cross at some $k$, at which $F_{0}\neq0$. To have Majoranas at $\omega_{1}/2$
quasienergy, we require, for example $\epsilon_{0}^{+}$ crosses $\epsilon_{p}^{-}$
at some $k$ when $F_{-p}\neq0$. Similarly, to have Majoranas at
$\omega_{2}/2$ quasienergy, we require, for example $\epsilon_{0}^{+}$
crosses $\epsilon_{q}^{-}$ at some $k$ when $F_{-q}\neq0$. Even
if we take static pairing $\Delta_{k}(t)=\Delta\sin k$, namely $\Delta_{k}^{(s)}=\delta_{s0}\Delta_{k}$
and $D_{r}=J_{r}^{p,q}(4\alpha_{1},4\alpha_{2})$, we can have both
types of Majoranas, for example taking nonvanishing $J_{0}^{p,q}(4\alpha_{1},4\alpha_{2}),J_{p}^{p,q}(4\alpha_{1},4\alpha_{2}),J_{q}^{p,q}(4\alpha_{1},4\alpha_{2})$.

\section*{Signatures of Majorana multiplexing in correlation functions}
\subsection*{Majorana operators in second quantization}
Before analyzing signature of Majorana multiplexing, it is helpful to first introduce Majorana
operators in second-quantization. Let $\ket{\Psi_\alpha (t)} =\exp(-i\epsilon_\alpha t) \ket{\Phi_\alpha (t)}$ be a solution to the time-dependent Schr\"odinger
equation 
\begin{equation}
i\partial_t \ket{\Psi_\alpha (t)} = H(t) \ket{\Psi_\alpha (t)}, 
\end{equation}
where  $H(t)$ and $\ket{\Phi_\alpha(t)}$ are time-quasiperiodic with the same frequencies, and $\epsilon_\alpha$ is the
quasienergy. Creation (annihilation) operators $\psi_\alpha^\dagger (t)$ ($\psi_\alpha (t) =
(\psi_\alpha^\dagger(t))^\dagger$) corresponding to $\ket{\Psi_\alpha (t)}$ can be defined as
\begin{equation}
\psi_\alpha^\dagger (t) = \sum_{j} C_j^\dagger \braket{j|\Psi_{\alpha}(t)}, 
\end{equation}
where $j$ is the real space index, $C_j^\dagger$ is the creation operator (may have multicomponents) at position $j$ and
$\braket{j|\Psi_{\alpha}(t)}$ is the real space wave function (with the same number of components as in $C_j$) of
$\ket{\Psi_\alpha (t)}$.

In the case of time-quasiperiodic Kitaev chain, we have
\begin{equation}
\psi_\alpha (t) = e^{-i\epsilon_\alpha t} \sum_{j=1}^N\left[ c_j^\dagger \phi_{\alpha,e}(j,t) + c_j\phi_{\alpha,h}(j,t)\right], 
\end{equation}
where $\phi_{\alpha,e}(j,t)$ and $\phi_{\alpha,h}(j,t)$ are the two components in
the Nambu wave function $\braket{j|\phi_\alpha(t)} =
(\phi_{\alpha,e}(j,t),\phi_{\alpha,h}(j,t))$. Due to time-quasiperiodicity, we have
\begin{equation}
\phi_{\alpha,e/h}(j,t) = \sum_{\mathbf{m}}\exp(-i\mathbf{m}\cdot \boldsymbol{\omega})
\phi_{\mathbf{m},e/h}^\alpha(j),
\end{equation} 
where $\phi_{\alpha,\mathbf{m},e/h}(j)$ are the solution of Eq.~(1) in the main text 
represented in both real space and the synthetic lattice.  
For Majorana operators, in particular, we have $\psi_{\alpha}(t) =
\psi_{\alpha}^\dagger(t)$ for all $t$. This restricts the quasienergy to be $\mathbf{n}\cdot\boldsymbol{\omega}/2$. 
The wave function at quasienergy $\epsilon_{\mathbf{n}} =\mathbf{n}\cdot\boldsymbol{\omega}/2 $ is also restricted
to satisfy $\phi_{\mathbf{m},h} = \phi_{-(\mathbf{m+n}),e}^*$.

When there are two frequencies $\omega_1$ and $\omega_2$, the Majorana operators  of the
chain at quasienergies $0$, $\omega_1/2$ and $\omega_2/2$ can be written as
\begin{equation}
\psi_0^\dagger (t) =  \sum_{j=1}^N \sum_{n,m} e^{-i(n\omega_1+m\omega_2)t}\left[ \phi_{n,m,e}(j)c_j^\dagger +
\phi_{n,m,h}(j) c_{j}\right] \simeq \sum_{j=1} \left[\phi_{0,0,e}(j) c_j^\dagger + \phi_{0,0,h}(j) c_j\right] ,
\end{equation}
\begin{align}
\psi_1^\dagger (t) &=\sum_{j=1}^N \sum_{n,m} e^{-i(n\omega_1+m\omega_2)t} 
\left[e^{i \omega_1 t/2} \phi_{n-1,m,e}(j)c_j^\dagger + e^{-i\omega_1 t/2} \phi_{n,m,h}(j) c_{j} \right]  \\
&\simeq  \sum_{j=1}^N \left[e^{i \omega_1 t/2} \phi_{-1,0,e}(j)c_j^\dagger + e^{-i\omega_1 t/2} \phi_{0,0,h}(j) c_{j}
\right] ,
\end{align}
and 
\begin{align}
\psi_2^\dagger (t) &=\sum_{j=1}^N \sum_{n,m} e^{-i(n\omega_1+m\omega_2)t} 
\left[e^{i \omega_2 t/2} \phi_{n,m,e}(j)c_j^\dagger + e^{-i\omega_2 t/2} \phi_{n,m-1,h}(j) c_{j} \right]  \\
&\simeq  \sum_{j=1}^N \left[e^{i \omega_2 t/2} \phi_{0,-1,e}(j)c_j^\dagger + e^{-i\omega_2 t/2} \phi_{0,0,h}(j) c_{j} \right] 
\end{align}
respectively. 
For Majoranas localized near the first site of the chain, the functions $\phi_{n,m,e/h}(j)$ appeared in the above
expressions decays exponentially as $j$ increases.

\subsection*{Correlation function}
The presense of Majoranas of different types at the end of a time-quasiperiodic Kitaev chain can be detected
using correlation functions of some local operators, such as the single particle Green's function. 
To be concrete, let us consider $\bra{0}\gamma_1(t)\gamma_1(0)\ket{0}$ where $\gamma_1 = (c_1 +
c_1^\dagger)/\sqrt{2}$, and $\ket{0}$ represents the BCS vaccuum at $t=0$.
The existence of Majoranas localized around the first site enables
us to write
\begin{equation}
\gamma_1(t) = c_0\psi_0(t) + c_1\psi_1(t) + c_2\psi_2(t) + \dots, 
\end{equation}
where $\dots$ includes other extended state which has less contribution compared to the Majoranas. 
Hence, we have $\bra{0}\gamma_1(t)\gamma_1(0)\ket{0}$ will oscillate at frequencies $\omega_1/2$ and $\omega_2/2$. 

\subsection*{Temporal disorder}
To explore the robustness of these Majoranas in the presense of temporal disorder, we consider
exponential correlated Gaussian noise in the drive. We replace $\omega_i t$ by $\omega_{i}t + \delta_i(t)$
with 
\begin{equation}
	\braket{\dot{\delta}_i(t)\dot{\delta}_j(t')} = \delta_{ij}\sigma^2 \exp(-\abs{t-t'}/\tau_d), 
	\label{eq:noise}
\end{equation}
where $\tau_d$ is the the correlation time, and $\delta_{i}(t)$ is a Gaussian distributed random variable with zero mean and variance $\sigma$.

\begin{figure}[t]
	\centering
	\includegraphics[width=0.45\textwidth]{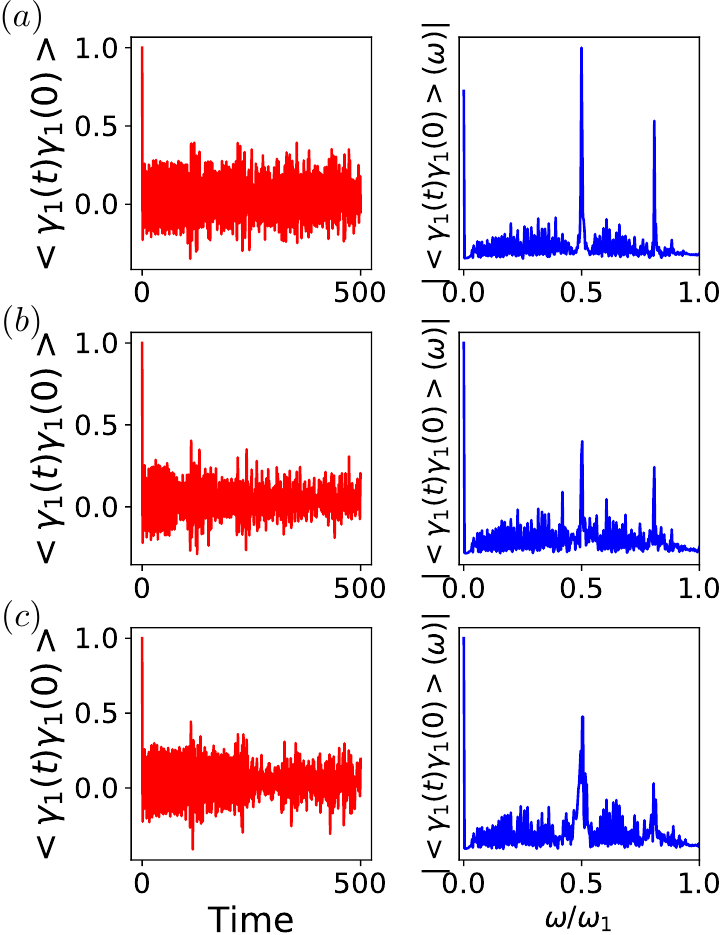}
	\caption{\label{sfig:disorder}}  Time evolution of $\braket{\gamma_1(t)\gamma_1(0)}$ (left panels)
	and its Fourier transform in the frequency domain (right panels), 
	simulated on the time-quasiperiodic Kitaev chain, with addtional correlated Gaussian noise defined in
  Eq.~(\ref{eq:noise}). 
	The other parameters are the same as in Fig.~4 of the main text.
	The parameters for the noise are  $\sigma = 0.1$;  $(a)$ $\omega_2 \tau_d = 1$, $(b)$ $\omega_2\tau_d = 20$, and $(c)$
  $\omega_2\tau_d = n100$. 
\end{figure}

In Fig.~\ref{sfig:disorder}, we show two numerical simulations of $\braket{\gamma_1(t)\gamma_1(0)}$ using the same parameters as the ones in the main text,
with additional correlated Gaussian noise. We see that peaks at $0$, $\omega_1/2$ and $\omega_2/2$ are robust against moderate disorder strength $\sigma$,
and correlation time $\tau_d$. As $\tau_d$ gets longer, these peaks get broader.

\subsection*{Commensurate frequencies}
Practically, the two frequencies $\omega_1$ and $\omega_2$ can hardly be mutually irrational. Let us assume $\omega_2/\omega_1 = p/q$, with $p,q\in \mathbb{Z}$. 
In the synthetic space,  the system is still Wannier-Stark localized along the electric field, while
perpendicular to the field it becomes periodic, with a large unit cell when $p$ and $q$ are large.
In this case, the wave functions are still localized within the unit cell due to the large
variation of on-site energies between different sites. We still have Majoranas from pariing within the same site or between neighboring sites. 

Let approximate the golden ratio $(\sqrt{2}+1)/2$ by $5/3$, and take $\omega_2/\omega_1=5/3$ for the time-dependent Kitaev chain. 
Fig.~\ref{sfig:commensurate} shows the wave function of the Majoranas in synthetic space and in real space. 
We find that the Majorana amplitudes are only localized with unit cells perpendicular to the direction of the electric field. 
In Fig.~\ref{sfig:correlation-5-3}, we show the correlation $\braket{\gamma_1(t)\gamma_1(0)}$, and also find peaks at $\omega_1/2$ and $\omega_2/2$.

\begin{figure}[t]
	\centering
	\includegraphics[width=0.45\textwidth]{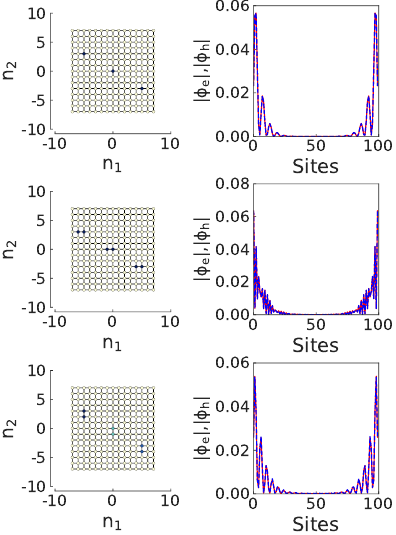}
	\caption{\label{sfig:commensurate} Numerical solution of the $0$-frequency and time-quasiperiodic Majorana states
		on the 2D synthetic lattice of size $15 \times 15$.
		Each site of the lattice corresponding to a Kitaev
chain of length $N = 100$. Left:  $\abs{\phi_{n_1,n_2}}^2$ for the $0$, $\frac{\omega_1}{2}$, and $\frac{\omega_2}{2}$  Majoranas
on the 2D synthetic lattice, where the darker color corresponds to a larger magnitude.
Right: the absolute value of the corresponding Majorana wave function, summed over the 2D synthetic lattice.
The electron and hole components $\phi_e, \phi_h$ are plotted as red solid and blue dashed curves.
The other parameters are $\omega_2/\omega_1 = 5/3$, $J/\omega_1=0.51$, $\mu/\omega_1=0.87$, $\Delta/\omega_1=0.051$,
and   $\Delta'/\omega_1=0.038$.}
\end{figure}

\begin{figure}[t]
	\centering
	\includegraphics[width=0.45\textwidth]{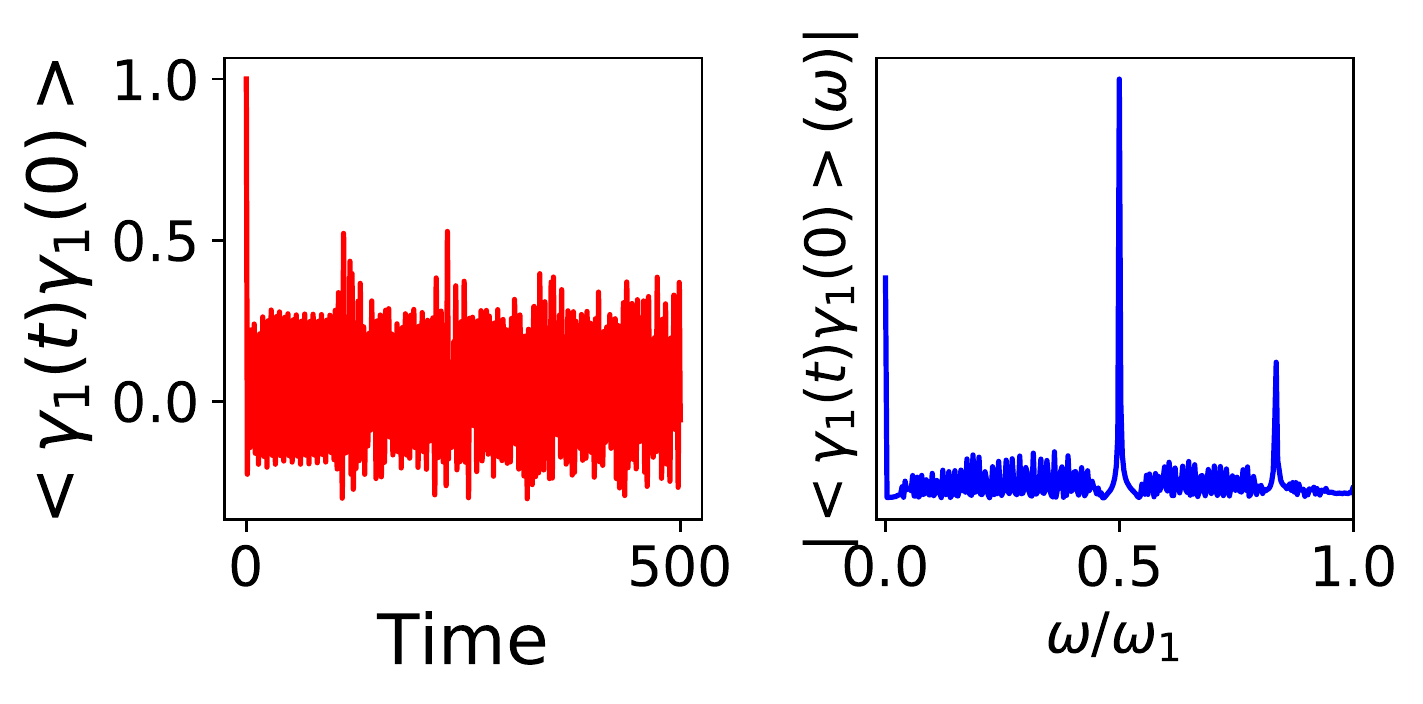}
	\caption{\label{sfig:correlation-5-3} Time evolution of $\braket{\gamma_1(t)\gamma_1(0)}$ (left panels)
		and its Fourier transform in the frequency domain (right panels), with  $\omega_2/\omega_1=5/3$,
		$J/\omega_1=0.51$, $\mu/\omega_1=0.87$, $\Delta/\omega_1=0.051$,
and   $\Delta'/\omega_1=0.038$. 	 }
\end{figure}

\end{widetext}

\begin{thebibliography}{57}%
\makeatletter
\providecommand \@ifxundefined [1]{%
 \@ifx{#1\undefined}
}%
\providecommand \@ifnum [1]{%
 \ifnum #1\expandafter \@firstoftwo
 \else \expandafter \@secondoftwo
 \fi
}%
\providecommand \@ifx [1]{%
 \ifx #1\expandafter \@firstoftwo
 \else \expandafter \@secondoftwo
 \fi
}%
\providecommand \natexlab [1]{#1}%
\providecommand \enquote  [1]{``#1''}%
\providecommand \bibnamefont  [1]{#1}%
\providecommand \bibfnamefont [1]{#1}%
\providecommand \citenamefont [1]{#1}%
\providecommand \href@noop [0]{\@secondoftwo}%
\providecommand \href [0]{\begingroup \@sanitize@url \@href}%
\providecommand \@href[1]{\@@startlink{#1}\@@href}%
\providecommand \@@href[1]{\endgroup#1\@@endlink}%
\providecommand \@sanitize@url [0]{\catcode `\\12\catcode `\$12\catcode
  `\&12\catcode `\#12\catcode `\^12\catcode `\_12\catcode `\%12\relax}%
\providecommand \@@startlink[1]{}%
\providecommand \@@endlink[0]{}%
\providecommand \url  [0]{\begingroup\@sanitize@url \@url }%
\providecommand \@url [1]{\endgroup\@href {#1}{\urlprefix }}%
\providecommand \urlprefix  [0]{URL }%
\providecommand \Eprint [0]{\href }%
\providecommand \doibase [0]{http://dx.doi.org/}%
\providecommand \selectlanguage [0]{\@gobble}%
\providecommand \bibinfo  [0]{\@secondoftwo}%
\providecommand \bibfield  [0]{\@secondoftwo}%
\providecommand \translation [1]{[#1]}%
\providecommand \BibitemOpen [0]{}%
\providecommand \bibitemStop [0]{}%
\providecommand \bibitemNoStop [0]{.\EOS\space}%
\providecommand \EOS [0]{\spacefactor3000\relax}%
\providecommand \BibitemShut  [1]{\csname bibitem#1\endcsname}%
\let\auto@bib@innerbib\@empty
%</preamble>
\bibitem [{\citenamefont {Kitaev}(2001)}]{Kitaev2001}%
  \BibitemOpen
  \bibfield  {author} {\bibinfo {author} {\bibfnamefont {A.}~\bibnamefont
  {Kitaev}},\ }\href {http://stacks.iop.org/1063-7869/44/i=10S/a=S29}
  {\bibfield  {journal} {\bibinfo  {journal} {Phys. Usp.}\ }\textbf {\bibinfo
  {volume} {44}},\ \bibinfo {pages} {131} (\bibinfo {year} {2001})}\BibitemShut
  {NoStop}%
\bibitem [{\citenamefont {Alicea}(2012)}]{Alicea2012}%
  \BibitemOpen
  \bibfield  {author} {\bibinfo {author} {\bibfnamefont {J.}~\bibnamefont
  {Alicea}},\ }\href@noop {} {\bibfield  {journal} {\bibinfo  {journal}
  {Reports on progress in physics}\ }\textbf {\bibinfo {volume} {75}},\
  \bibinfo {pages} {076501} (\bibinfo {year} {2012})}\BibitemShut {NoStop}%
\bibitem [{\citenamefont {Beenakker}(2013)}]{Beenakker2013}%
  \BibitemOpen
  \bibfield  {author} {\bibinfo {author} {\bibfnamefont {C.}~\bibnamefont
  {Beenakker}},\ }\href@noop {} {\  (\bibinfo {year} {2013})}\BibitemShut
  {NoStop}%
\bibitem [{\citenamefont {Kitaev}(2003)}]{Kitaev2003}%
  \BibitemOpen
  \bibfield  {author} {\bibinfo {author} {\bibfnamefont {A.~Y.}\ \bibnamefont
  {Kitaev}},\ }\href {\doibase 10.1016/s0003-4916(02)00018-0} {\bibfield
  {journal} {\bibinfo  {journal} {Ann. Phys.}\ }\textbf {\bibinfo {volume}
  {303}},\ \bibinfo {pages} {2} (\bibinfo {year} {2003})}\BibitemShut {NoStop}%
\bibitem [{\citenamefont {Nayak}\ \emph {et~al.}(2008)\citenamefont {Nayak},
  \citenamefont {Simon}, \citenamefont {Stern}, \citenamefont {Freedman},\ and\
  \citenamefont {Das~Sarma}}]{Nayak2008}%
  \BibitemOpen
  \bibfield  {author} {\bibinfo {author} {\bibfnamefont {C.}~\bibnamefont
  {Nayak}}, \bibinfo {author} {\bibfnamefont {S.~H.}\ \bibnamefont {Simon}},
  \bibinfo {author} {\bibfnamefont {A.}~\bibnamefont {Stern}}, \bibinfo
  {author} {\bibfnamefont {M.}~\bibnamefont {Freedman}}, \ and\ \bibinfo
  {author} {\bibfnamefont {S.}~\bibnamefont {Das~Sarma}},\ }\href {\doibase
  10.1103/RevModPhys.80.1083} {\bibfield  {journal} {\bibinfo  {journal} {Rev.
  Mod. Phys.}\ }\textbf {\bibinfo {volume} {80}},\ \bibinfo {pages} {1083}
  (\bibinfo {year} {2008})}\BibitemShut {NoStop}%
\bibitem [{\citenamefont {Aasen}\ \emph {et~al.}(2016)\citenamefont {Aasen},
  \citenamefont {Hell}, \citenamefont {Mishmash}, \citenamefont {Higginbotham},
  \citenamefont {Danon}, \citenamefont {Leijnse}, \citenamefont {Jespersen},
  \citenamefont {Folk}, \citenamefont {Marcus}, \citenamefont {Flensberg},\
  and\ \citenamefont {Alicea}}]{Aasen2016}%
  \BibitemOpen
  \bibfield  {author} {\bibinfo {author} {\bibfnamefont {D.}~\bibnamefont
  {Aasen}}, \bibinfo {author} {\bibfnamefont {M.}~\bibnamefont {Hell}},
  \bibinfo {author} {\bibfnamefont {R.~V.}\ \bibnamefont {Mishmash}}, \bibinfo
  {author} {\bibfnamefont {A.}~\bibnamefont {Higginbotham}}, \bibinfo {author}
  {\bibfnamefont {J.}~\bibnamefont {Danon}}, \bibinfo {author} {\bibfnamefont
  {M.}~\bibnamefont {Leijnse}}, \bibinfo {author} {\bibfnamefont {T.~S.}\
  \bibnamefont {Jespersen}}, \bibinfo {author} {\bibfnamefont {J.~A.}\
  \bibnamefont {Folk}}, \bibinfo {author} {\bibfnamefont {C.~M.}\ \bibnamefont
  {Marcus}}, \bibinfo {author} {\bibfnamefont {K.}~\bibnamefont {Flensberg}}, \
  and\ \bibinfo {author} {\bibfnamefont {J.}~\bibnamefont {Alicea}},\ }\href
  {\doibase 10.1103/PhysRevX.6.031016} {\bibfield  {journal} {\bibinfo
  {journal} {Phys. Rev. X}\ }\textbf {\bibinfo {volume} {6}},\ \bibinfo {pages}
  {031016} (\bibinfo {year} {2016})}\BibitemShut {NoStop}%
\bibitem [{\citenamefont {Fu}\ and\ \citenamefont {Kane}(2008)}]{Fu2008}%
  \BibitemOpen
  \bibfield  {author} {\bibinfo {author} {\bibfnamefont {L.}~\bibnamefont
  {Fu}}\ and\ \bibinfo {author} {\bibfnamefont {C.~L.}\ \bibnamefont {Kane}},\
  }\href {\doibase 10.1103/PhysRevLett.100.096407} {\bibfield  {journal}
  {\bibinfo  {journal} {Phys. Rev. Lett.}\ }\textbf {\bibinfo {volume} {100}},\
  \bibinfo {pages} {096407} (\bibinfo {year} {2008})}\BibitemShut {NoStop}%
\bibitem [{\citenamefont {Zhang}\ \emph {et~al.}(2008)\citenamefont {Zhang},
  \citenamefont {Tewari}, \citenamefont {Lutchyn},\ and\ \citenamefont
  {Das~Sarma}}]{Zhang2008}%
  \BibitemOpen
  \bibfield  {author} {\bibinfo {author} {\bibfnamefont {C.}~\bibnamefont
  {Zhang}}, \bibinfo {author} {\bibfnamefont {S.}~\bibnamefont {Tewari}},
  \bibinfo {author} {\bibfnamefont {R.~M.}\ \bibnamefont {Lutchyn}}, \ and\
  \bibinfo {author} {\bibfnamefont {S.}~\bibnamefont {Das~Sarma}},\ }\href
  {\doibase 10.1103/PhysRevLett.101.160401} {\bibfield  {journal} {\bibinfo
  {journal} {Phys. Rev. Lett.}\ }\textbf {\bibinfo {volume} {101}},\ \bibinfo
  {pages} {160401} (\bibinfo {year} {2008})}\BibitemShut {NoStop}%
\bibitem [{\citenamefont {Sato}\ \emph {et~al.}(2009)\citenamefont {Sato},
  \citenamefont {Takahashi},\ and\ \citenamefont {Fujimoto}}]{Sato2009}%
  \BibitemOpen
  \bibfield  {author} {\bibinfo {author} {\bibfnamefont {M.}~\bibnamefont
  {Sato}}, \bibinfo {author} {\bibfnamefont {Y.}~\bibnamefont {Takahashi}}, \
  and\ \bibinfo {author} {\bibfnamefont {S.}~\bibnamefont {Fujimoto}},\ }\href
  {\doibase 10.1103/PhysRevLett.103.020401} {\bibfield  {journal} {\bibinfo
  {journal} {Phys. Rev. Lett.}\ }\textbf {\bibinfo {volume} {103}},\ \bibinfo
  {pages} {020401} (\bibinfo {year} {2009})}\BibitemShut {NoStop}%
\bibitem [{\citenamefont {Lutchyn}\ \emph {et~al.}(2010)\citenamefont
  {Lutchyn}, \citenamefont {Sau},\ and\ \citenamefont
  {Das~Sarma}}]{Lutchyn2010}%
  \BibitemOpen
  \bibfield  {author} {\bibinfo {author} {\bibfnamefont {R.~M.}\ \bibnamefont
  {Lutchyn}}, \bibinfo {author} {\bibfnamefont {J.~D.}\ \bibnamefont {Sau}}, \
  and\ \bibinfo {author} {\bibfnamefont {S.}~\bibnamefont {Das~Sarma}},\ }\href
  {\doibase 10.1103/PhysRevLett.105.077001} {\bibfield  {journal} {\bibinfo
  {journal} {Phys. Rev. Lett.}\ }\textbf {\bibinfo {volume} {105}},\ \bibinfo
  {pages} {077001} (\bibinfo {year} {2010})}\BibitemShut {NoStop}%
\bibitem [{\citenamefont {Oreg}\ \emph {et~al.}(2010)\citenamefont {Oreg},
  \citenamefont {Refael},\ and\ \citenamefont {von Oppen}}]{Oreg2010}%
  \BibitemOpen
  \bibfield  {author} {\bibinfo {author} {\bibfnamefont {Y.}~\bibnamefont
  {Oreg}}, \bibinfo {author} {\bibfnamefont {G.}~\bibnamefont {Refael}}, \ and\
  \bibinfo {author} {\bibfnamefont {F.}~\bibnamefont {von Oppen}},\ }\href
  {\doibase 10.1103/PhysRevLett.105.177002} {\bibfield  {journal} {\bibinfo
  {journal} {Phys. Rev. Lett.}\ }\textbf {\bibinfo {volume} {105}},\ \bibinfo
  {pages} {177002} (\bibinfo {year} {2010})}\BibitemShut {NoStop}%
\bibitem [{\citenamefont {Diehl}\ \emph {et~al.}(2011)\citenamefont {Diehl},
  \citenamefont {Rico}, \citenamefont {Baranov},\ and\ \citenamefont
  {Zoller}}]{Diehl2011}%
  \BibitemOpen
  \bibfield  {author} {\bibinfo {author} {\bibfnamefont {S.}~\bibnamefont
  {Diehl}}, \bibinfo {author} {\bibfnamefont {E.}~\bibnamefont {Rico}},
  \bibinfo {author} {\bibfnamefont {M.~A.}\ \bibnamefont {Baranov}}, \ and\
  \bibinfo {author} {\bibfnamefont {P.}~\bibnamefont {Zoller}},\ }\href@noop {}
  {\bibfield  {journal} {\bibinfo  {journal} {Nature Physics}\ }\textbf
  {\bibinfo {volume} {7}},\ \bibinfo {pages} {971} (\bibinfo {year}
  {2011})}\BibitemShut {NoStop}%
\bibitem [{\citenamefont {Jiang}\ \emph {et~al.}(2011)\citenamefont {Jiang},
  \citenamefont {Kitagawa}, \citenamefont {Alicea}, \citenamefont {Akhmerov},
  \citenamefont {Pekker}, \citenamefont {Refael}, \citenamefont {Cirac},
  \citenamefont {Demler}, \citenamefont {Lukin},\ and\ \citenamefont
  {Zoller}}]{Jiang2011}%
  \BibitemOpen
  \bibfield  {author} {\bibinfo {author} {\bibfnamefont {L.}~\bibnamefont
  {Jiang}}, \bibinfo {author} {\bibfnamefont {T.}~\bibnamefont {Kitagawa}},
  \bibinfo {author} {\bibfnamefont {J.}~\bibnamefont {Alicea}}, \bibinfo
  {author} {\bibfnamefont {A.~R.}\ \bibnamefont {Akhmerov}}, \bibinfo {author}
  {\bibfnamefont {D.}~\bibnamefont {Pekker}}, \bibinfo {author} {\bibfnamefont
  {G.}~\bibnamefont {Refael}}, \bibinfo {author} {\bibfnamefont {J.~I.}\
  \bibnamefont {Cirac}}, \bibinfo {author} {\bibfnamefont {E.}~\bibnamefont
  {Demler}}, \bibinfo {author} {\bibfnamefont {M.~D.}\ \bibnamefont {Lukin}}, \
  and\ \bibinfo {author} {\bibfnamefont {P.}~\bibnamefont {Zoller}},\ }\href
  {\doibase 10.1103/PhysRevLett.106.220402} {\bibfield  {journal} {\bibinfo
  {journal} {Phys. Rev. Lett.}\ }\textbf {\bibinfo {volume} {106}},\ \bibinfo
  {pages} {220402} (\bibinfo {year} {2011})}\BibitemShut {NoStop}%
\bibitem [{\citenamefont {Nadj-Perge}\ \emph {et~al.}(2013)\citenamefont
  {Nadj-Perge}, \citenamefont {Drozdov}, \citenamefont {Bernevig},\ and\
  \citenamefont {Yazdani}}]{Nadj-Perge2013}%
  \BibitemOpen
  \bibfield  {author} {\bibinfo {author} {\bibfnamefont {S.}~\bibnamefont
  {Nadj-Perge}}, \bibinfo {author} {\bibfnamefont {I.~K.}\ \bibnamefont
  {Drozdov}}, \bibinfo {author} {\bibfnamefont {B.~A.}\ \bibnamefont
  {Bernevig}}, \ and\ \bibinfo {author} {\bibfnamefont {A.}~\bibnamefont
  {Yazdani}},\ }\href {\doibase 10.1103/PhysRevB.88.020407} {\bibfield
  {journal} {\bibinfo  {journal} {Phys. Rev. B}\ }\textbf {\bibinfo {volume}
  {88}},\ \bibinfo {pages} {020407} (\bibinfo {year} {2013})}\BibitemShut
  {NoStop}%
\bibitem [{\citenamefont {Pientka}\ \emph {et~al.}(2013)\citenamefont
  {Pientka}, \citenamefont {Glazman},\ and\ \citenamefont {von
  Oppen}}]{Pientka2013}%
  \BibitemOpen
  \bibfield  {author} {\bibinfo {author} {\bibfnamefont {F.}~\bibnamefont
  {Pientka}}, \bibinfo {author} {\bibfnamefont {L.~I.}\ \bibnamefont
  {Glazman}}, \ and\ \bibinfo {author} {\bibfnamefont {F.}~\bibnamefont {von
  Oppen}},\ }\href {\doibase 10.1103/PhysRevB.88.155420} {\bibfield  {journal}
  {\bibinfo  {journal} {Phys. Rev. B}\ }\textbf {\bibinfo {volume} {88}},\
  \bibinfo {pages} {155420} (\bibinfo {year} {2013})}\BibitemShut {NoStop}%
\bibitem [{\citenamefont {Foster}\ \emph {et~al.}(2014)\citenamefont {Foster},
  \citenamefont {Gurarie}, \citenamefont {Dzero},\ and\ \citenamefont
  {Yuzbashyan}}]{Foster2014}%
  \BibitemOpen
  \bibfield  {author} {\bibinfo {author} {\bibfnamefont {M.~S.}\ \bibnamefont
  {Foster}}, \bibinfo {author} {\bibfnamefont {V.}~\bibnamefont {Gurarie}},
  \bibinfo {author} {\bibfnamefont {M.}~\bibnamefont {Dzero}}, \ and\ \bibinfo
  {author} {\bibfnamefont {E.~A.}\ \bibnamefont {Yuzbashyan}},\ }\href
  {\doibase 10.1103/PhysRevLett.113.076403} {\bibfield  {journal} {\bibinfo
  {journal} {Phys. Rev. Lett.}\ }\textbf {\bibinfo {volume} {113}},\ \bibinfo
  {pages} {076403} (\bibinfo {year} {2014})}\BibitemShut {NoStop}%
\bibitem [{\citenamefont {Peng}\ \emph {et~al.}(2015)\citenamefont {Peng},
  \citenamefont {Pientka}, \citenamefont {Glazman},\ and\ \citenamefont {von
  Oppen}}]{Peng2015}%
  \BibitemOpen
  \bibfield  {author} {\bibinfo {author} {\bibfnamefont {Y.}~\bibnamefont
  {Peng}}, \bibinfo {author} {\bibfnamefont {F.}~\bibnamefont {Pientka}},
  \bibinfo {author} {\bibfnamefont {L.~I.}\ \bibnamefont {Glazman}}, \ and\
  \bibinfo {author} {\bibfnamefont {F.}~\bibnamefont {von Oppen}},\ }\href
  {\doibase 10.1103/PhysRevLett.114.106801} {\bibfield  {journal} {\bibinfo
  {journal} {Phys. Rev. Lett.}\ }\textbf {\bibinfo {volume} {114}},\ \bibinfo
  {pages} {106801} (\bibinfo {year} {2015})}\BibitemShut {NoStop}%
\bibitem [{\citenamefont {Mourik}\ \emph {et~al.}(2012)\citenamefont {Mourik},
  \citenamefont {Zuo}, \citenamefont {Frolov}, \citenamefont {Plissard},
  \citenamefont {Bakkers},\ and\ \citenamefont {Kouwenhoven}}]{Mourik2012}%
  \BibitemOpen
  \bibfield  {author} {\bibinfo {author} {\bibfnamefont {V.}~\bibnamefont
  {Mourik}}, \bibinfo {author} {\bibfnamefont {K.}~\bibnamefont {Zuo}},
  \bibinfo {author} {\bibfnamefont {S.~M.}\ \bibnamefont {Frolov}}, \bibinfo
  {author} {\bibfnamefont {S.}~\bibnamefont {Plissard}}, \bibinfo {author}
  {\bibfnamefont {E.~P.}\ \bibnamefont {Bakkers}}, \ and\ \bibinfo {author}
  {\bibfnamefont {L.~P.}\ \bibnamefont {Kouwenhoven}},\ }\href@noop {}
  {\bibfield  {journal} {\bibinfo  {journal} {Science}\ }\textbf {\bibinfo
  {volume} {336}},\ \bibinfo {pages} {1003} (\bibinfo {year}
  {2012})}\BibitemShut {NoStop}%
\bibitem [{\citenamefont {Das}\ \emph {et~al.}(2012)\citenamefont {Das},
  \citenamefont {Ronen}, \citenamefont {Most}, \citenamefont {Oreg},
  \citenamefont {Heiblum},\ and\ \citenamefont {Shtrikman}}]{Das2012}%
  \BibitemOpen
  \bibfield  {author} {\bibinfo {author} {\bibfnamefont {A.}~\bibnamefont
  {Das}}, \bibinfo {author} {\bibfnamefont {Y.}~\bibnamefont {Ronen}}, \bibinfo
  {author} {\bibfnamefont {Y.}~\bibnamefont {Most}}, \bibinfo {author}
  {\bibfnamefont {Y.}~\bibnamefont {Oreg}}, \bibinfo {author} {\bibfnamefont
  {M.}~\bibnamefont {Heiblum}}, \ and\ \bibinfo {author} {\bibfnamefont
  {H.}~\bibnamefont {Shtrikman}},\ }\href@noop {} {\bibfield  {journal}
  {\bibinfo  {journal} {Nature Physics}\ }\textbf {\bibinfo {volume} {8}},\
  \bibinfo {pages} {887} (\bibinfo {year} {2012})}\BibitemShut {NoStop}%
\bibitem [{\citenamefont {Churchill}\ \emph {et~al.}(2013)\citenamefont
  {Churchill}, \citenamefont {Fatemi}, \citenamefont {Grove-Rasmussen},
  \citenamefont {Deng}, \citenamefont {Caroff}, \citenamefont {Xu},\ and\
  \citenamefont {Marcus}}]{Churchill2013}%
  \BibitemOpen
  \bibfield  {author} {\bibinfo {author} {\bibfnamefont {H.~O.~H.}\
  \bibnamefont {Churchill}}, \bibinfo {author} {\bibfnamefont {V.}~\bibnamefont
  {Fatemi}}, \bibinfo {author} {\bibfnamefont {K.}~\bibnamefont
  {Grove-Rasmussen}}, \bibinfo {author} {\bibfnamefont {M.~T.}\ \bibnamefont
  {Deng}}, \bibinfo {author} {\bibfnamefont {P.}~\bibnamefont {Caroff}},
  \bibinfo {author} {\bibfnamefont {H.~Q.}\ \bibnamefont {Xu}}, \ and\ \bibinfo
  {author} {\bibfnamefont {C.~M.}\ \bibnamefont {Marcus}},\ }\href {\doibase
  10.1103/PhysRevB.87.241401} {\bibfield  {journal} {\bibinfo  {journal} {Phys.
  Rev. B}\ }\textbf {\bibinfo {volume} {87}},\ \bibinfo {pages} {241401}
  (\bibinfo {year} {2013})}\BibitemShut {NoStop}%
\bibitem [{\citenamefont {Deng}\ \emph {et~al.}(2012)\citenamefont {Deng},
  \citenamefont {Yu}, \citenamefont {Huang}, \citenamefont {Larsson},
  \citenamefont {Caroff},\ and\ \citenamefont {Xu}}]{Deng2012}%
  \BibitemOpen
  \bibfield  {author} {\bibinfo {author} {\bibfnamefont {M.}~\bibnamefont
  {Deng}}, \bibinfo {author} {\bibfnamefont {C.}~\bibnamefont {Yu}}, \bibinfo
  {author} {\bibfnamefont {G.}~\bibnamefont {Huang}}, \bibinfo {author}
  {\bibfnamefont {M.}~\bibnamefont {Larsson}}, \bibinfo {author} {\bibfnamefont
  {P.}~\bibnamefont {Caroff}}, \ and\ \bibinfo {author} {\bibfnamefont
  {H.}~\bibnamefont {Xu}},\ }\href@noop {} {\bibfield  {journal} {\bibinfo
  {journal} {Nano letters}\ }\textbf {\bibinfo {volume} {12}},\ \bibinfo
  {pages} {6414} (\bibinfo {year} {2012})}\BibitemShut {NoStop}%
\bibitem [{\citenamefont {Finck}\ \emph {et~al.}(2013)\citenamefont {Finck},
  \citenamefont {Van~Harlingen}, \citenamefont {Mohseni}, \citenamefont
  {Jung},\ and\ \citenamefont {Li}}]{Finck2013}%
  \BibitemOpen
  \bibfield  {author} {\bibinfo {author} {\bibfnamefont {A.}~\bibnamefont
  {Finck}}, \bibinfo {author} {\bibfnamefont {D.}~\bibnamefont
  {Van~Harlingen}}, \bibinfo {author} {\bibfnamefont {P.}~\bibnamefont
  {Mohseni}}, \bibinfo {author} {\bibfnamefont {K.}~\bibnamefont {Jung}}, \
  and\ \bibinfo {author} {\bibfnamefont {X.}~\bibnamefont {Li}},\ }\href@noop
  {} {\bibfield  {journal} {\bibinfo  {journal} {Physical review letters}\
  }\textbf {\bibinfo {volume} {110}},\ \bibinfo {pages} {126406} (\bibinfo
  {year} {2013})}\BibitemShut {NoStop}%
\bibitem [{\citenamefont {Nadj-Perge}\ \emph {et~al.}(2014)\citenamefont
  {Nadj-Perge}, \citenamefont {Drozdov}, \citenamefont {Li}, \citenamefont
  {Chen}, \citenamefont {Jeon}, \citenamefont {Seo}, \citenamefont {MacDonald},
  \citenamefont {Bernevig},\ and\ \citenamefont {Yazdani}}]{Nadj-Perge2014}%
  \BibitemOpen
  \bibfield  {author} {\bibinfo {author} {\bibfnamefont {S.}~\bibnamefont
  {Nadj-Perge}}, \bibinfo {author} {\bibfnamefont {I.~K.}\ \bibnamefont
  {Drozdov}}, \bibinfo {author} {\bibfnamefont {J.}~\bibnamefont {Li}},
  \bibinfo {author} {\bibfnamefont {H.}~\bibnamefont {Chen}}, \bibinfo {author}
  {\bibfnamefont {S.}~\bibnamefont {Jeon}}, \bibinfo {author} {\bibfnamefont
  {J.}~\bibnamefont {Seo}}, \bibinfo {author} {\bibfnamefont {A.~H.}\
  \bibnamefont {MacDonald}}, \bibinfo {author} {\bibfnamefont {B.~A.}\
  \bibnamefont {Bernevig}}, \ and\ \bibinfo {author} {\bibfnamefont
  {A.}~\bibnamefont {Yazdani}},\ }\href@noop {} {\bibfield  {journal} {\bibinfo
   {journal} {Science}\ }\textbf {\bibinfo {volume} {346}},\ \bibinfo {pages}
  {602} (\bibinfo {year} {2014})}\BibitemShut {NoStop}%
\bibitem [{\citenamefont {Ruby}\ \emph {et~al.}(2015)\citenamefont {Ruby},
  \citenamefont {Pientka}, \citenamefont {Peng}, \citenamefont {von Oppen},
  \citenamefont {Heinrich},\ and\ \citenamefont {Franke}}]{Ruby2015}%
  \BibitemOpen
  \bibfield  {author} {\bibinfo {author} {\bibfnamefont {M.}~\bibnamefont
  {Ruby}}, \bibinfo {author} {\bibfnamefont {F.}~\bibnamefont {Pientka}},
  \bibinfo {author} {\bibfnamefont {Y.}~\bibnamefont {Peng}}, \bibinfo {author}
  {\bibfnamefont {F.}~\bibnamefont {von Oppen}}, \bibinfo {author}
  {\bibfnamefont {B.~W.}\ \bibnamefont {Heinrich}}, \ and\ \bibinfo {author}
  {\bibfnamefont {K.~J.}\ \bibnamefont {Franke}},\ }\href {\doibase
  10.1103/PhysRevLett.115.197204} {\bibfield  {journal} {\bibinfo  {journal}
  {Phys. Rev. Lett.}\ }\textbf {\bibinfo {volume} {115}},\ \bibinfo {pages}
  {197204} (\bibinfo {year} {2015})}\BibitemShut {NoStop}%
\bibitem [{\citenamefont {Pawlak}\ \emph {et~al.}(2016)\citenamefont {Pawlak},
  \citenamefont {Kisiel}, \citenamefont {Klinovaja}, \citenamefont {Meier},
  \citenamefont {Kawai}, \citenamefont {Glatzel}, \citenamefont {Loss},\ and\
  \citenamefont {Meyer}}]{Pawlak2016}%
  \BibitemOpen
  \bibfield  {author} {\bibinfo {author} {\bibfnamefont {R.}~\bibnamefont
  {Pawlak}}, \bibinfo {author} {\bibfnamefont {M.}~\bibnamefont {Kisiel}},
  \bibinfo {author} {\bibfnamefont {J.}~\bibnamefont {Klinovaja}}, \bibinfo
  {author} {\bibfnamefont {T.}~\bibnamefont {Meier}}, \bibinfo {author}
  {\bibfnamefont {S.}~\bibnamefont {Kawai}}, \bibinfo {author} {\bibfnamefont
  {T.}~\bibnamefont {Glatzel}}, \bibinfo {author} {\bibfnamefont
  {D.}~\bibnamefont {Loss}}, \ and\ \bibinfo {author} {\bibfnamefont
  {E.}~\bibnamefont {Meyer}},\ }\href@noop {} {\bibfield  {journal} {\bibinfo
  {journal} {npj Quantum Information}\ }\textbf {\bibinfo {volume} {2}},\
  \bibinfo {pages} {16035} (\bibinfo {year} {2016})}\BibitemShut {NoStop}%
\bibitem [{\citenamefont {Deng}\ \emph {et~al.}(2016)\citenamefont {Deng},
  \citenamefont {Vaitiek{\.e}nas}, \citenamefont {Hansen}, \citenamefont
  {Danon}, \citenamefont {Leijnse}, \citenamefont {Flensberg}, \citenamefont
  {Nyg{\aa}rd}, \citenamefont {Krogstrup},\ and\ \citenamefont
  {Marcus}}]{Deng2016}%
  \BibitemOpen
  \bibfield  {author} {\bibinfo {author} {\bibfnamefont {M.}~\bibnamefont
  {Deng}}, \bibinfo {author} {\bibfnamefont {S.}~\bibnamefont
  {Vaitiek{\.e}nas}}, \bibinfo {author} {\bibfnamefont {E.~B.}\ \bibnamefont
  {Hansen}}, \bibinfo {author} {\bibfnamefont {J.}~\bibnamefont {Danon}},
  \bibinfo {author} {\bibfnamefont {M.}~\bibnamefont {Leijnse}}, \bibinfo
  {author} {\bibfnamefont {K.}~\bibnamefont {Flensberg}}, \bibinfo {author}
  {\bibfnamefont {J.}~\bibnamefont {Nyg{\aa}rd}}, \bibinfo {author}
  {\bibfnamefont {P.}~\bibnamefont {Krogstrup}}, \ and\ \bibinfo {author}
  {\bibfnamefont {C.~M.}\ \bibnamefont {Marcus}},\ }\href@noop {} {\bibfield
  {journal} {\bibinfo  {journal} {Science}\ }\textbf {\bibinfo {volume}
  {354}},\ \bibinfo {pages} {1557} (\bibinfo {year} {2016})}\BibitemShut
  {NoStop}%
\bibitem [{\citenamefont {Albrecht}\ \emph {et~al.}(2016)\citenamefont
  {Albrecht}, \citenamefont {Higginbotham}, \citenamefont {Madsen},
  \citenamefont {Kuemmeth}, \citenamefont {Jespersen}, \citenamefont
  {Nyg{\aa}rd}, \citenamefont {Krogstrup},\ and\ \citenamefont
  {Marcus}}]{Albrecht2016}%
  \BibitemOpen
  \bibfield  {author} {\bibinfo {author} {\bibfnamefont {S.~M.}\ \bibnamefont
  {Albrecht}}, \bibinfo {author} {\bibfnamefont {A.}~\bibnamefont
  {Higginbotham}}, \bibinfo {author} {\bibfnamefont {M.}~\bibnamefont
  {Madsen}}, \bibinfo {author} {\bibfnamefont {F.}~\bibnamefont {Kuemmeth}},
  \bibinfo {author} {\bibfnamefont {T.~S.}\ \bibnamefont {Jespersen}}, \bibinfo
  {author} {\bibfnamefont {J.}~\bibnamefont {Nyg{\aa}rd}}, \bibinfo {author}
  {\bibfnamefont {P.}~\bibnamefont {Krogstrup}}, \ and\ \bibinfo {author}
  {\bibfnamefont {C.}~\bibnamefont {Marcus}},\ }\href@noop {} {\bibfield
  {journal} {\bibinfo  {journal} {Nature}\ }\textbf {\bibinfo {volume} {531}},\
  \bibinfo {pages} {206} (\bibinfo {year} {2016})}\BibitemShut {NoStop}%
\bibitem [{\citenamefont {Ruby}\ \emph {et~al.}(2017)\citenamefont {Ruby},
  \citenamefont {Heinrich}, \citenamefont {Peng}, \citenamefont {von Oppen},\
  and\ \citenamefont {Franke}}]{Ruby2017}%
  \BibitemOpen
  \bibfield  {author} {\bibinfo {author} {\bibfnamefont {M.}~\bibnamefont
  {Ruby}}, \bibinfo {author} {\bibfnamefont {B.~W.}\ \bibnamefont {Heinrich}},
  \bibinfo {author} {\bibfnamefont {Y.}~\bibnamefont {Peng}}, \bibinfo {author}
  {\bibfnamefont {F.}~\bibnamefont {von Oppen}}, \ and\ \bibinfo {author}
  {\bibfnamefont {K.~J.}\ \bibnamefont {Franke}},\ }\href@noop {} {\bibfield
  {journal} {\bibinfo  {journal} {Nano letters}\ }\textbf {\bibinfo {volume}
  {17}},\ \bibinfo {pages} {4473} (\bibinfo {year} {2017})}\BibitemShut
  {NoStop}%
\bibitem [{\citenamefont {G{\"u}l}\ \emph {et~al.}(2018)\citenamefont
  {G{\"u}l}, \citenamefont {Zhang}, \citenamefont {Bommer}, \citenamefont
  {de~Moor}, \citenamefont {Car}, \citenamefont {Plissard}, \citenamefont
  {Bakkers}, \citenamefont {Geresdi}, \citenamefont {Watanabe}, \citenamefont
  {Taniguchi} \emph {et~al.}}]{Gul2018}%
  \BibitemOpen
  \bibfield  {author} {\bibinfo {author} {\bibfnamefont {{\"O}.}~\bibnamefont
  {G{\"u}l}}, \bibinfo {author} {\bibfnamefont {H.}~\bibnamefont {Zhang}},
  \bibinfo {author} {\bibfnamefont {J.~D.}\ \bibnamefont {Bommer}}, \bibinfo
  {author} {\bibfnamefont {M.~W.}\ \bibnamefont {de~Moor}}, \bibinfo {author}
  {\bibfnamefont {D.}~\bibnamefont {Car}}, \bibinfo {author} {\bibfnamefont
  {S.~R.}\ \bibnamefont {Plissard}}, \bibinfo {author} {\bibfnamefont {E.~P.}\
  \bibnamefont {Bakkers}}, \bibinfo {author} {\bibfnamefont {A.}~\bibnamefont
  {Geresdi}}, \bibinfo {author} {\bibfnamefont {K.}~\bibnamefont {Watanabe}},
  \bibinfo {author} {\bibfnamefont {T.}~\bibnamefont {Taniguchi}},  \emph
  {et~al.},\ }\href@noop {} {\bibfield  {journal} {\bibinfo  {journal} {Nature
  nanotechnology}\ ,\ \bibinfo {pages} {1}} (\bibinfo {year}
  {2018})}\BibitemShut {NoStop}%
\bibitem [{\citenamefont {Klinovaja}\ \emph {et~al.}(2016)\citenamefont
  {Klinovaja}, \citenamefont {Stano},\ and\ \citenamefont
  {Loss}}]{Klinovaja2016}%
  \BibitemOpen
  \bibfield  {author} {\bibinfo {author} {\bibfnamefont {J.}~\bibnamefont
  {Klinovaja}}, \bibinfo {author} {\bibfnamefont {P.}~\bibnamefont {Stano}}, \
  and\ \bibinfo {author} {\bibfnamefont {D.}~\bibnamefont {Loss}},\ }\href
  {\doibase 10.1103/PhysRevLett.116.176401} {\bibfield  {journal} {\bibinfo
  {journal} {Phys. Rev. Lett.}\ }\textbf {\bibinfo {volume} {116}},\ \bibinfo
  {pages} {176401} (\bibinfo {year} {2016})}\BibitemShut {NoStop}%
\bibitem [{\citenamefont {Thakurathi}\ \emph {et~al.}(2017)\citenamefont
  {Thakurathi}, \citenamefont {Loss},\ and\ \citenamefont
  {Klinovaja}}]{Thakurathi2017}%
  \BibitemOpen
  \bibfield  {author} {\bibinfo {author} {\bibfnamefont {M.}~\bibnamefont
  {Thakurathi}}, \bibinfo {author} {\bibfnamefont {D.}~\bibnamefont {Loss}}, \
  and\ \bibinfo {author} {\bibfnamefont {J.}~\bibnamefont {Klinovaja}},\ }\href
  {\doibase 10.1103/PhysRevB.95.155407} {\bibfield  {journal} {\bibinfo
  {journal} {Phys. Rev. B}\ }\textbf {\bibinfo {volume} {95}},\ \bibinfo
  {pages} {155407} (\bibinfo {year} {2017})}\BibitemShut {NoStop}%
\bibitem [{\citenamefont {Wang}\ \emph {et~al.}(2013)\citenamefont {Wang},
  \citenamefont {Steinberg}, \citenamefont {Jarillo-Herrero},\ and\
  \citenamefont {Gedik}}]{Wang2013}%
  \BibitemOpen
  \bibfield  {author} {\bibinfo {author} {\bibfnamefont {Y.}~\bibnamefont
  {Wang}}, \bibinfo {author} {\bibfnamefont {H.}~\bibnamefont {Steinberg}},
  \bibinfo {author} {\bibfnamefont {P.}~\bibnamefont {Jarillo-Herrero}}, \ and\
  \bibinfo {author} {\bibfnamefont {N.}~\bibnamefont {Gedik}},\ }\href@noop {}
  {\bibfield  {journal} {\bibinfo  {journal} {Science}\ }\textbf {\bibinfo
  {volume} {342}},\ \bibinfo {pages} {453} (\bibinfo {year}
  {2013})}\BibitemShut {NoStop}%
\bibitem [{\citenamefont {Jotzu}\ \emph {et~al.}(2014)\citenamefont {Jotzu},
  \citenamefont {Messer}, \citenamefont {Desbuquois}, \citenamefont {Lebrat},
  \citenamefont {Uehlinger}, \citenamefont {Greif},\ and\ \citenamefont
  {Esslinger}}]{Jotzu2014}%
  \BibitemOpen
  \bibfield  {author} {\bibinfo {author} {\bibfnamefont {G.}~\bibnamefont
  {Jotzu}}, \bibinfo {author} {\bibfnamefont {M.}~\bibnamefont {Messer}},
  \bibinfo {author} {\bibfnamefont {R.}~\bibnamefont {Desbuquois}}, \bibinfo
  {author} {\bibfnamefont {M.}~\bibnamefont {Lebrat}}, \bibinfo {author}
  {\bibfnamefont {T.}~\bibnamefont {Uehlinger}}, \bibinfo {author}
  {\bibfnamefont {D.}~\bibnamefont {Greif}}, \ and\ \bibinfo {author}
  {\bibfnamefont {T.}~\bibnamefont {Esslinger}},\ }\href@noop {} {\bibfield
  {journal} {\bibinfo  {journal} {Nature}\ }\textbf {\bibinfo {volume} {515}},\
  \bibinfo {pages} {237} (\bibinfo {year} {2014})}\BibitemShut {NoStop}%
\bibitem [{\citenamefont {Aidelsburger}\ \emph {et~al.}(2015)\citenamefont
  {Aidelsburger}, \citenamefont {Lohse}, \citenamefont {Schweizer},
  \citenamefont {Atala}, \citenamefont {Barreiro}, \citenamefont {Nascimbene},
  \citenamefont {Cooper}, \citenamefont {Bloch},\ and\ \citenamefont
  {Goldman}}]{Aidelsburger2015}%
  \BibitemOpen
  \bibfield  {author} {\bibinfo {author} {\bibfnamefont {M.}~\bibnamefont
  {Aidelsburger}}, \bibinfo {author} {\bibfnamefont {M.}~\bibnamefont {Lohse}},
  \bibinfo {author} {\bibfnamefont {C.}~\bibnamefont {Schweizer}}, \bibinfo
  {author} {\bibfnamefont {M.}~\bibnamefont {Atala}}, \bibinfo {author}
  {\bibfnamefont {J.~T.}\ \bibnamefont {Barreiro}}, \bibinfo {author}
  {\bibfnamefont {S.}~\bibnamefont {Nascimbene}}, \bibinfo {author}
  {\bibfnamefont {N.}~\bibnamefont {Cooper}}, \bibinfo {author} {\bibfnamefont
  {I.}~\bibnamefont {Bloch}}, \ and\ \bibinfo {author} {\bibfnamefont
  {N.}~\bibnamefont {Goldman}},\ }\href@noop {} {\bibfield  {journal} {\bibinfo
   {journal} {Nature Physics}\ }\textbf {\bibinfo {volume} {11}},\ \bibinfo
  {pages} {162} (\bibinfo {year} {2015})}\BibitemShut {NoStop}%
\bibitem [{\citenamefont {Tarnowski}\ \emph {et~al.}(2017)\citenamefont
  {Tarnowski}, \citenamefont {{\"U}nal}, \citenamefont {Fl{\"a}schner},
  \citenamefont {Rem}, \citenamefont {Eckardt}, \citenamefont {Sengstock},\
  and\ \citenamefont {Weitenberg}}]{Tarnowski2017}%
  \BibitemOpen
  \bibfield  {author} {\bibinfo {author} {\bibfnamefont {M.}~\bibnamefont
  {Tarnowski}}, \bibinfo {author} {\bibfnamefont {F.~N.}\ \bibnamefont
  {{\"U}nal}}, \bibinfo {author} {\bibfnamefont {N.}~\bibnamefont
  {Fl{\"a}schner}}, \bibinfo {author} {\bibfnamefont {B.~S.}\ \bibnamefont
  {Rem}}, \bibinfo {author} {\bibfnamefont {A.}~\bibnamefont {Eckardt}},
  \bibinfo {author} {\bibfnamefont {K.}~\bibnamefont {Sengstock}}, \ and\
  \bibinfo {author} {\bibfnamefont {C.}~\bibnamefont {Weitenberg}},\
  }\href@noop {} {\bibfield  {journal} {\bibinfo  {journal} {arXiv:1709.01046}\
  } (\bibinfo {year} {2017})}\BibitemShut {NoStop}%
\bibitem [{\citenamefont {Maczewsky}\ \emph {et~al.}(2017)\citenamefont
  {Maczewsky}, \citenamefont {Zeuner}, \citenamefont {Nolte},\ and\
  \citenamefont {Szameit}}]{Maczewsky2017}%
  \BibitemOpen
  \bibfield  {author} {\bibinfo {author} {\bibfnamefont {L.~J.}\ \bibnamefont
  {Maczewsky}}, \bibinfo {author} {\bibfnamefont {J.~M.}\ \bibnamefont
  {Zeuner}}, \bibinfo {author} {\bibfnamefont {S.}~\bibnamefont {Nolte}}, \
  and\ \bibinfo {author} {\bibfnamefont {A.}~\bibnamefont {Szameit}},\
  }\href@noop {} {\bibfield  {journal} {\bibinfo  {journal} {Nature
  communications}\ }\textbf {\bibinfo {volume} {8}},\ \bibinfo {pages} {13756}
  (\bibinfo {year} {2017})}\BibitemShut {NoStop}%
\bibitem [{\citenamefont {Liu}\ \emph {et~al.}(2013)\citenamefont {Liu},
  \citenamefont {Levchenko},\ and\ \citenamefont {Baranger}}]{Liu2013}%
  \BibitemOpen
  \bibfield  {author} {\bibinfo {author} {\bibfnamefont {D.~E.}\ \bibnamefont
  {Liu}}, \bibinfo {author} {\bibfnamefont {A.}~\bibnamefont {Levchenko}}, \
  and\ \bibinfo {author} {\bibfnamefont {H.~U.}\ \bibnamefont {Baranger}},\
  }\href {\doibase 10.1103/PhysRevLett.111.047002} {\bibfield  {journal}
  {\bibinfo  {journal} {Phys. Rev. Lett.}\ }\textbf {\bibinfo {volume} {111}},\
  \bibinfo {pages} {047002} (\bibinfo {year} {2013})}\BibitemShut {NoStop}%
\bibitem [{\citenamefont {Karzig}\ \emph {et~al.}()\citenamefont {Karzig},
  \citenamefont {Bauer}, \citenamefont {Pereg-Barnea}, \citenamefont {Oreg},\
  and\ \citenamefont {Refael}}]{time-braiding}%
  \BibitemOpen
  \bibfield  {author} {\bibinfo {author} {\bibfnamefont {T.}~\bibnamefont
  {Karzig}}, \bibinfo {author} {\bibfnamefont {B.}~\bibnamefont {Bauer}},
  \bibinfo {author} {\bibfnamefont {T.}~\bibnamefont {Pereg-Barnea}}, \bibinfo
  {author} {\bibfnamefont {Y.}~\bibnamefont {Oreg}}, \ and\ \bibinfo {author}
  {\bibfnamefont {G.}~\bibnamefont {Refael}},\ }\href@noop {} {}\bibinfo
  {howpublished} {in preparation}\BibitemShut {NoStop}%
\bibitem [{\citenamefont {Tomlinson}(1977)}]{Tomlinson1977}%
  \BibitemOpen
  \bibfield  {author} {\bibinfo {author} {\bibfnamefont {W.}~\bibnamefont
  {Tomlinson}},\ }\href@noop {} {\bibfield  {journal} {\bibinfo  {journal}
  {Applied Optics}\ }\textbf {\bibinfo {volume} {16}},\ \bibinfo {pages} {2180}
  (\bibinfo {year} {1977})}\BibitemShut {NoStop}%
\bibitem [{\citenamefont {Dumitrescu}\ \emph {et~al.}(2018)\citenamefont
  {Dumitrescu}, \citenamefont {Vasseur},\ and\ \citenamefont
  {Potter}}]{Dumitrescu2018}%
  \BibitemOpen
  \bibfield  {author} {\bibinfo {author} {\bibfnamefont {P.~T.}\ \bibnamefont
  {Dumitrescu}}, \bibinfo {author} {\bibfnamefont {R.}~\bibnamefont {Vasseur}},
  \ and\ \bibinfo {author} {\bibfnamefont {A.~C.}\ \bibnamefont {Potter}},\
  }\href {\doibase 10.1103/PhysRevLett.120.070602} {\bibfield  {journal}
  {\bibinfo  {journal} {Phys. Rev. Lett.}\ }\textbf {\bibinfo {volume} {120}},\
  \bibinfo {pages} {070602} (\bibinfo {year} {2018})}\BibitemShut {NoStop}%
\bibitem [{\citenamefont {Li}\ \emph {et~al.}(2012)\citenamefont {Li},
  \citenamefont {Gong}, \citenamefont {Yin}, \citenamefont {Quan},
  \citenamefont {Yin}, \citenamefont {Zhang}, \citenamefont {Duan},\ and\
  \citenamefont {Zhang}}]{Li2012}%
  \BibitemOpen
  \bibfield  {author} {\bibinfo {author} {\bibfnamefont {T.}~\bibnamefont
  {Li}}, \bibinfo {author} {\bibfnamefont {Z.-X.}\ \bibnamefont {Gong}},
  \bibinfo {author} {\bibfnamefont {Z.-Q.}\ \bibnamefont {Yin}}, \bibinfo
  {author} {\bibfnamefont {H.~T.}\ \bibnamefont {Quan}}, \bibinfo {author}
  {\bibfnamefont {X.}~\bibnamefont {Yin}}, \bibinfo {author} {\bibfnamefont
  {P.}~\bibnamefont {Zhang}}, \bibinfo {author} {\bibfnamefont {L.-M.}\
  \bibnamefont {Duan}}, \ and\ \bibinfo {author} {\bibfnamefont
  {X.}~\bibnamefont {Zhang}},\ }\href {\doibase 10.1103/PhysRevLett.109.163001}
  {\bibfield  {journal} {\bibinfo  {journal} {Phys. Rev. Lett.}\ }\textbf
  {\bibinfo {volume} {109}},\ \bibinfo {pages} {163001} (\bibinfo {year}
  {2012})}\BibitemShut {NoStop}%
\bibitem [{\citenamefont {Flicker}(2017)}]{Flicker2017}%
  \BibitemOpen
  \bibfield  {author} {\bibinfo {author} {\bibfnamefont {F.}~\bibnamefont
  {Flicker}},\ }\href@noop {} {\bibfield  {journal} {\bibinfo  {journal} {arXiv
  preprint arXiv:1707.09371}\ } (\bibinfo {year} {2017})}\BibitemShut {NoStop}%
\bibitem [{\citenamefont {Autti}\ \emph {et~al.}(2018)\citenamefont {Autti},
  \citenamefont {Eltsov},\ and\ \citenamefont {Volovik}}]{Autti2018}%
  \BibitemOpen
  \bibfield  {author} {\bibinfo {author} {\bibfnamefont {S.}~\bibnamefont
  {Autti}}, \bibinfo {author} {\bibfnamefont {V.~B.}\ \bibnamefont {Eltsov}}, \
  and\ \bibinfo {author} {\bibfnamefont {G.~E.}\ \bibnamefont {Volovik}},\
  }\href {\doibase 10.1103/PhysRevLett.120.215301} {\bibfield  {journal}
  {\bibinfo  {journal} {Phys. Rev. Lett.}\ }\textbf {\bibinfo {volume} {120}},\
  \bibinfo {pages} {215301} (\bibinfo {year} {2018})}\BibitemShut {NoStop}%
\bibitem [{\citenamefont {Huang}\ \emph {et~al.}(2018)\citenamefont {Huang},
  \citenamefont {Li},\ and\ \citenamefont {Yin}}]{Huang2018}%
  \BibitemOpen
  \bibfield  {author} {\bibinfo {author} {\bibfnamefont {Y.}~\bibnamefont
  {Huang}}, \bibinfo {author} {\bibfnamefont {T.}~\bibnamefont {Li}}, \ and\
  \bibinfo {author} {\bibfnamefont {Z.-q.}\ \bibnamefont {Yin}},\ }\href
  {\doibase 10.1103/PhysRevA.97.012115} {\bibfield  {journal} {\bibinfo
  {journal} {Phys. Rev. A}\ }\textbf {\bibinfo {volume} {97}},\ \bibinfo
  {pages} {012115} (\bibinfo {year} {2018})}\BibitemShut {NoStop}%
\bibitem [{\citenamefont {Giergiel}\ \emph {et~al.}(2018)\citenamefont
  {Giergiel}, \citenamefont {Miroszewski},\ and\ \citenamefont
  {Sacha}}]{Giergiel2018}%
  \BibitemOpen
  \bibfield  {author} {\bibinfo {author} {\bibfnamefont {K.}~\bibnamefont
  {Giergiel}}, \bibinfo {author} {\bibfnamefont {A.}~\bibnamefont
  {Miroszewski}}, \ and\ \bibinfo {author} {\bibfnamefont {K.}~\bibnamefont
  {Sacha}},\ }\href {\doibase 10.1103/PhysRevLett.120.140401} {\bibfield
  {journal} {\bibinfo  {journal} {Phys. Rev. Lett.}\ }\textbf {\bibinfo
  {volume} {120}},\ \bibinfo {pages} {140401} (\bibinfo {year}
  {2018})}\BibitemShut {NoStop}%
\bibitem [{\citenamefont {Khemani}\ \emph {et~al.}(2016)\citenamefont
  {Khemani}, \citenamefont {Lazarides}, \citenamefont {Moessner},\ and\
  \citenamefont {Sondhi}}]{Khemani2016}%
  \BibitemOpen
  \bibfield  {author} {\bibinfo {author} {\bibfnamefont {V.}~\bibnamefont
  {Khemani}}, \bibinfo {author} {\bibfnamefont {A.}~\bibnamefont {Lazarides}},
  \bibinfo {author} {\bibfnamefont {R.}~\bibnamefont {Moessner}}, \ and\
  \bibinfo {author} {\bibfnamefont {S.~L.}\ \bibnamefont {Sondhi}},\ }\href
  {\doibase 10.1103/PhysRevLett.116.250401} {\bibfield  {journal} {\bibinfo
  {journal} {Phys. Rev. Lett.}\ }\textbf {\bibinfo {volume} {116}},\ \bibinfo
  {pages} {250401} (\bibinfo {year} {2016})}\BibitemShut {NoStop}%
\bibitem [{\citenamefont {Else}\ \emph {et~al.}(2016)\citenamefont {Else},
  \citenamefont {Bauer},\ and\ \citenamefont {Nayak}}]{Else2016}%
  \BibitemOpen
  \bibfield  {author} {\bibinfo {author} {\bibfnamefont {D.~V.}\ \bibnamefont
  {Else}}, \bibinfo {author} {\bibfnamefont {B.}~\bibnamefont {Bauer}}, \ and\
  \bibinfo {author} {\bibfnamefont {C.}~\bibnamefont {Nayak}},\ }\href
  {\doibase 10.1103/PhysRevLett.117.090402} {\bibfield  {journal} {\bibinfo
  {journal} {Phys. Rev. Lett.}\ }\textbf {\bibinfo {volume} {117}},\ \bibinfo
  {pages} {090402} (\bibinfo {year} {2016})}\BibitemShut {NoStop}%
\bibitem [{\citenamefont {Potter}\ \emph {et~al.}(2016)\citenamefont {Potter},
  \citenamefont {Morimoto},\ and\ \citenamefont {Vishwanath}}]{Potter2016}%
  \BibitemOpen
  \bibfield  {author} {\bibinfo {author} {\bibfnamefont {A.~C.}\ \bibnamefont
  {Potter}}, \bibinfo {author} {\bibfnamefont {T.}~\bibnamefont {Morimoto}}, \
  and\ \bibinfo {author} {\bibfnamefont {A.}~\bibnamefont {Vishwanath}},\
  }\href {\doibase 10.1103/PhysRevX.6.041001} {\bibfield  {journal} {\bibinfo
  {journal} {Phys. Rev. X}\ }\textbf {\bibinfo {volume} {6}},\ \bibinfo {pages}
  {041001} (\bibinfo {year} {2016})}\BibitemShut {NoStop}%
\bibitem [{\citenamefont {Bomantara}\ and\ \citenamefont
  {Gong}(2018)}]{Bomantara2018}%
  \BibitemOpen
  \bibfield  {author} {\bibinfo {author} {\bibfnamefont {R.~W.}\ \bibnamefont
  {Bomantara}}\ and\ \bibinfo {author} {\bibfnamefont {J.}~\bibnamefont
  {Gong}},\ }\href {\doibase 10.1103/PhysRevLett.120.230405} {\bibfield
  {journal} {\bibinfo  {journal} {Phys. Rev. Lett.}\ }\textbf {\bibinfo
  {volume} {120}},\ \bibinfo {pages} {230405} (\bibinfo {year}
  {2018})}\BibitemShut {NoStop}%
\bibitem [{\citenamefont {Fukuyama}\ \emph {et~al.}(1973)\citenamefont
  {Fukuyama}, \citenamefont {Bari},\ and\ \citenamefont
  {Fogedby}}]{Fukuyama1973}%
  \BibitemOpen
  \bibfield  {author} {\bibinfo {author} {\bibfnamefont {H.}~\bibnamefont
  {Fukuyama}}, \bibinfo {author} {\bibfnamefont {R.~A.}\ \bibnamefont {Bari}},
  \ and\ \bibinfo {author} {\bibfnamefont {H.~C.}\ \bibnamefont {Fogedby}},\
  }\href {\doibase 10.1103/PhysRevB.8.5579} {\bibfield  {journal} {\bibinfo
  {journal} {Phys. Rev. B}\ }\textbf {\bibinfo {volume} {8}},\ \bibinfo {pages}
  {5579} (\bibinfo {year} {1973})}\BibitemShut {NoStop}%
\bibitem [{\citenamefont {Emin}\ and\ \citenamefont {Hart}(1987)}]{Emin1987}%
  \BibitemOpen
  \bibfield  {author} {\bibinfo {author} {\bibfnamefont {D.}~\bibnamefont
  {Emin}}\ and\ \bibinfo {author} {\bibfnamefont {C.~F.}\ \bibnamefont
  {Hart}},\ }\href {\doibase 10.1103/PhysRevB.36.7353} {\bibfield  {journal}
  {\bibinfo  {journal} {Phys. Rev. B}\ }\textbf {\bibinfo {volume} {36}},\
  \bibinfo {pages} {7353} (\bibinfo {year} {1987})}\BibitemShut {NoStop}%
\bibitem [{sup()}]{suppl}%
  \BibitemOpen
  \href@noop {} {\bibinfo  {journal} {Supplemental Material}\ }\BibitemShut
  {NoStop}%
\bibitem [{\citenamefont {Martin}\ \emph {et~al.}(2017)\citenamefont {Martin},
  \citenamefont {Refael},\ and\ \citenamefont {Halperin}}]{Martin2017}%
  \BibitemOpen
\bibfield  {journal} {  }\bibfield  {author} {\bibinfo {author} {\bibfnamefont
  {I.}~\bibnamefont {Martin}}, \bibinfo {author} {\bibfnamefont
  {G.}~\bibnamefont {Refael}}, \ and\ \bibinfo {author} {\bibfnamefont
  {B.}~\bibnamefont {Halperin}},\ }\href {\doibase 10.1103/PhysRevX.7.041008}
  {\bibfield  {journal} {\bibinfo  {journal} {Phys. Rev. X}\ }\textbf {\bibinfo
  {volume} {7}},\ \bibinfo {pages} {041008} (\bibinfo {year}
  {2017})}\BibitemShut {NoStop}%
\bibitem [{\citenamefont {Peng}\ and\ \citenamefont {Refael}(2018)}]{Peng2018}%
  \BibitemOpen
  \bibfield  {author} {\bibinfo {author} {\bibfnamefont {Y.}~\bibnamefont
  {Peng}}\ and\ \bibinfo {author} {\bibfnamefont {G.}~\bibnamefont {Refael}},\
  }\href {\doibase 10.1103/PhysRevB.97.134303} {\bibfield  {journal} {\bibinfo
  {journal} {Phys. Rev. B}\ }\textbf {\bibinfo {volume} {97}},\ \bibinfo
  {pages} {134303} (\bibinfo {year} {2018})}\BibitemShut {NoStop}%
\bibitem [{\citenamefont {Duneau}\ and\ \citenamefont
  {Katz}(1985)}]{Duneau1985}%
  \BibitemOpen
  \bibfield  {author} {\bibinfo {author} {\bibfnamefont {M.}~\bibnamefont
  {Duneau}}\ and\ \bibinfo {author} {\bibfnamefont {A.}~\bibnamefont {Katz}},\
  }\href {\doibase 10.1103/PhysRevLett.54.2688} {\bibfield  {journal} {\bibinfo
   {journal} {Phys. Rev. Lett.}\ }\textbf {\bibinfo {volume} {54}},\ \bibinfo
  {pages} {2688} (\bibinfo {year} {1985})}\BibitemShut {NoStop}%
\bibitem [{\citenamefont {Aubry}\ and\ \citenamefont
  {Andr{\'e}}(1980)}]{Aubry1980}%
  \BibitemOpen
  \bibfield  {author} {\bibinfo {author} {\bibfnamefont {S.}~\bibnamefont
  {Aubry}}\ and\ \bibinfo {author} {\bibfnamefont {G.}~\bibnamefont
  {Andr{\'e}}},\ }\href@noop {} {\bibfield  {journal} {\bibinfo  {journal}
  {Ann. Israel Phys. Soc}\ }\textbf {\bibinfo {volume} {3}},\ \bibinfo {pages}
  {18} (\bibinfo {year} {1980})}\BibitemShut {NoStop}%
\bibitem [{\citenamefont {Lahini}\ \emph {et~al.}(2009)\citenamefont {Lahini},
  \citenamefont {Pugatch}, \citenamefont {Pozzi}, \citenamefont {Sorel},
  \citenamefont {Morandotti}, \citenamefont {Davidson},\ and\ \citenamefont
  {Silberberg}}]{Lahini2009}%
  \BibitemOpen
  \bibfield  {author} {\bibinfo {author} {\bibfnamefont {Y.}~\bibnamefont
  {Lahini}}, \bibinfo {author} {\bibfnamefont {R.}~\bibnamefont {Pugatch}},
  \bibinfo {author} {\bibfnamefont {F.}~\bibnamefont {Pozzi}}, \bibinfo
  {author} {\bibfnamefont {M.}~\bibnamefont {Sorel}}, \bibinfo {author}
  {\bibfnamefont {R.}~\bibnamefont {Morandotti}}, \bibinfo {author}
  {\bibfnamefont {N.}~\bibnamefont {Davidson}}, \ and\ \bibinfo {author}
  {\bibfnamefont {Y.}~\bibnamefont {Silberberg}},\ }\href {\doibase
  10.1103/PhysRevLett.103.013901} {\bibfield  {journal} {\bibinfo  {journal}
  {Phys. Rev. Lett.}\ }\textbf {\bibinfo {volume} {103}},\ \bibinfo {pages}
  {013901} (\bibinfo {year} {2009})}\BibitemShut {NoStop}%
\end{thebibliography}
\end{document}